\PassOptionsToPackage{hyperfootnotes=false}{hyperref}  
\documentclass[twocolumn]{simplejournal}
\usepackage{booktabs}  
\usepackage{multirow}  

\title{A thermodynamic correction layer for machine learning interatomic potentials}

\author{
  Sarath Menon\orcidlink{0000-0002-6776-1213}$^{1}$\email{sarath.menon@ruhr-uni-bochum.de},
  Braulio S\'{a}nchez Ortega\orcidlink{0009-0002-7373-5019}$^{1}$,
  Ralf Drautz\orcidlink{0000-0001-7101-8804}$^{1}$\\[6pt]
  \affil{$^1$ICAMS, Ruhr University Bochum, 44801 Bochum, Germany}\\
  }

\date{\today}

\begin{document}

\begin{abstract}
Machine learning interatomic potentials enable atomistic simulations with near \textit{ab initio} accuracy over large length and time scales.
Foundation models extend this capability across the periodic table but inherit the thermodynamic description of the electronic structure methods used to generate their training data.
Small errors in relative phase stability can therefore produce large errors in transformation temperatures.
We introduce an additive thermodynamic correction layer that shifts phase free energies towards an assessed reference without retraining or modifying the base potential.
A first order thermodynamic construction converts the required phase free energy shifts directly into configuration energy targets, which are fitted using a linear atomic cluster expansion and refined iteratively from measured free energies.
We apply the method to melting and solid to solid transformations across elements with different crystal structures using a foundation GRACE model.
After the correction, all transformation temperatures lie within the prescribed tolerance of the assessed values, including Ca, Ti, and Zr, where a solid to solid transition and melting are corrected together.
Static energy curves, bulk moduli, and liquid structure remain close to those of the base model.
Vacancy formation energies, thermal expansion, and liquid density depend on the corrected phase balance and change where the correction is large.
The method requires neither retraining, differentiable dynamics, nor access to the parameters of the base potential.
It provides a modular route for combining transferable atomistic models with assessed phase thermodynamics.
\end{abstract}

\maketitle

\section{Introduction}

Machine learning interatomic potentials (MLIPs) enable atomistic simulations with near \textit{ab initio} accuracy at a fraction of the computational cost \cite{Behler2016,Deringer2019,Drautz2019,Unke2021}.
Equivariant and graph based models have improved accuracy and transferability \cite{Batzner2022,Batatia2022,Bochkarev2024}, while foundation models extend this capability across the periodic table \cite{Chen2022,Deng2023,Yang2024,Batatia2025,Lysogorskiy2026Foundation}.
These models enable simulations over length and time scales that are inaccessible to first principles calculations.

Finite temperature thermodynamics provides a demanding test for interatomic potentials.
Melting and solid to solid transformation temperatures are determined by crossings between phase free energies.
The relevant free energy differences are often only a few meV per atom, so small errors in relative phase stability can shift a transformation temperature by hundreds of kelvin.
Established methods allow phase free energies and transformation temperatures to be calculated from atomistic potentials \cite{Freitas2016,deKoning1999,Chew2023,Menon2021,Menon2024}.
The accuracy of these predictions therefore depends on the thermodynamics represented by the underlying potential.

Differences from experimental thermodynamics do not necessarily originate from insufficient training data or limitations of the MLIP itself.
The thermodynamic description learned by an MLIP reflects the electronic structure method used to generate its training data.
Density functional approximations can introduce systematic differences in cohesive properties and relative phase stability \cite{Csonka2009,Tran2016,Hinuma2017,Zhang2018SolidStability}.
These differences are also reflected in predicted melting temperatures and phase diagrams \cite{LiFang2017,LiFang2020,Forslund2025,Menon2026}.
For foundation models trained across many elements and phases, the resulting model provides a transferable representation of this electronic structure reference rather than an empirical description of experimental phase thermodynamics.

Several approaches have been used for the thermodynamic refinement of interaction models.
Transition temperatures have been used directly in the construction of classical interaction models, including aluminium \cite{Sturgeon2000}, the TIP4P/Ice water model \cite{Abascal2005}, and Ti \cite{Mendelev2016}.
Calculated transition temperatures have also been corrected using the linear relation between changes in relative phase free energy and the corresponding shift in transformation temperature \cite{Alfe2002}.
Thermodynamic perturbation theory has been used to correct melting points of machine learning force fields to their first principles reference \cite{Jinnouchi2019} and to compute free energy differences between descriptions with training only at the end points \cite{Jinnouchi2020}.
Gibbs free energies have also been introduced directly as targets in potential fitting \cite{Wei2026}.
Swinburne, Lapointe, and Marinica developed a model agnostic free energy estimator for potentials that are linear in their descriptors and used its parameter gradients to move the $\alpha$ to $\gamma$ transformation temperature of a nonmagnetic Fe model \cite{Swinburne2025}.
Recent approaches obtain thermodynamic gradients by differentiating through simulations or by trajectory reweighting \cite{Thaler2021,Fuchs2026,Jin2026}.
Such approaches generally require differentiable dynamics, access to the trainable model, or repeated optimization of its parameters.

Here, we introduce an additive thermodynamic correction layer that operates on a given interatomic potential.
An additive pair correction to a fixed machine learning potential has previously been fitted to experimental liquid structure data \cite{Matin2024}, and a Gaussian pair correction fitted to formation enthalpies of an assessed thermodynamic database has been added to the training data of a potential trained on r2SCAN data \cite{Zhong2026}.
Transformation temperatures have been targeted by adjusting coefficients of a DFT based polynomial potential for Si to the experimental melting temperature and its pressure dependence \cite{Bauerhenne2021}, the parameters of a linear descriptor model \cite{Swinburne2025}, and those of a message passing potential through differentiable trajectory reweighting \cite{Fuchs2026}.
In these cases the refined parameters are those of the potential itself.
In contrast, the base potential is kept fixed here and can have any functional form, and the transformation temperatures are imposed through a separate additive correction.
The distinguishing feature of our approach is that assessed phase thermodynamics is converted directly into configuration energy labels for a compact additive potential, without modifying or retraining the underlying model.
Because the base potential enters only through energies and sampled configurations, the procedure is independent of its functional form and can in principle be applied to classical potentials, specialized MLIPs, and foundation models.

We apply the method to melting and solid to solid transitions across a set of elements using a foundation GRACE model \cite{Bochkarev2024,Lysogorskiy2026Foundation,Bochkarev2026} as the unmodified base potential.
After the correction, the transformation temperatures of the full test set lie within the prescribed tolerance of their assessed values, including systems with competing solid phases.
We further show that the correction retains the main structural behavior of the base model.
Static energy curves, bulk moduli, and liquid structure remain close to those of the original potential.
Vacancy formation energies, thermal expansion, and liquid density are not included in the fit.
They change where the correction is large and in most cases move towards experiment.
The correction can also incorporate transformation temperatures at finite pressure, allowing pressure temperature phase boundaries to be corrected within the same layer.
The correction layer separates thermodynamic refinement from model construction and can be applied to existing foundation models without retraining them.

\section{Results}

\subsection{Thermodynamic correction as a learnable potential layer}
\label{sec:correction}

We first formulate how macroscopic thermodynamic information can be translated into an atomistic correction potential.

We consider a transition between phases $\alpha$ and $\beta$.
Here $\alpha$ denotes the phase that is stable below the transformation temperature and $\beta$ the phase that is stable above it.
For melting, $\alpha$ is the solid and $\beta$ the liquid.
Throughout, lower case letters denote quantities per atom, with $g$, $u$, and $s$ the Gibbs free energy, internal energy, and entropy of a phase.
For clarity, the derivation is written at zero pressure, where $g=u-Ts$.
At finite pressure, $u$ is replaced by the enthalpy $h=u+pv$ and the same relations carry over.

We compare two descriptions of the same material.
The assessed reference is denoted by $A$, with transformation temperature $T_A$.
The base potential is denoted by $B$, with transformation temperature $T_B$.
Superscripts $A$ and $B$ mark quantities of the two descriptions where required.

For a transition between phases $\alpha$ and $\beta$, we define the phase free energy difference as
$\Delta g_{\alpha\beta}(T)=g_{\alpha}(T)-g_{\beta}(T)$,
with $\Delta u_{\alpha\beta}(T)$ and $\Delta s_{\alpha\beta}(T)$ defined analogously.
The transformation temperatures follow from the crossings $\Delta g_{\alpha\beta}^{A}(T_A)=0$ and $\Delta g_{\alpha\beta}^{B}(T_B)=0$.

The correction must supply the missing free energy difference of the base potential at the reference transformation temperature.
The residual can be constructed in three ways, depending on the thermodynamic information available.

First, the residual can be obtained directly from the measured free energy branches of the base potential,

\begin{equation}
r_{\alpha\beta}
=
-
\Delta g_{\alpha\beta}^{B}(T_A).
\label{eq:targetdirect}
\end{equation}

Equation~\eqref{eq:targetdirect} makes no approximation but requires the phase free energy branches of the base potential at temperature $T_A$.

Second, the residual can be estimated from the crossing of the base potential.
Expanding Eq.~\eqref{eq:targetdirect} to first order around $T_B$ (Sec.~\ref{sec:fediff}) and using the crossing relation $\Delta u_{\alpha\beta}^{B}(T_B)=T_B\Delta s_{\alpha\beta}^{B}(T_B)$ gives

\begin{equation}
r_{\alpha\beta}
\simeq
\Delta s_{\alpha\beta}^{B}(T_B)
\left(
T_A
-
T_B
\right)
=
\Delta u_{\alpha\beta}^{B}(T_B)
\frac{
T_A
-
T_B
}{
T_B
}.
\label{eq:targetB}
\end{equation}

Equation~\eqref{eq:targetB} requires the measured transformation temperature and transition entropy of the current potential at temperature $T_B$, together with the reference transformation temperature $T_A$.
In this work, $\Delta s_{\alpha\beta}$ is taken from the slope of the measured free energy difference at the crossing.

Third, if the assessed transition enthalpy is available, the residual can be estimated from phase energy differences at $T_A$.
If the transition entropies of both descriptions are assumed to coincide at $T_A$, the residual equals the difference between the assessed transition enthalpy and the phase energy difference of the base potential at $T_A$, Eq.~\eqref{eq:targetA} in Sec.~\ref{sec:fediff}.
This construction does not require a free energy calculation for the base potential, but it is sensitive to differences between the assessed and base transition entropies.

The residual is translated into an atomistic correction through the corrected potential $U_C=U_B+\Delta U$,
where $\Delta U$ is the additive correction.
To first order in $\Delta U$, thermodynamic perturbation theory gives the change in the phase free energy as the ensemble average of the correction energy.
The required phase difference is therefore

\begin{equation}
r_{\alpha\beta}
\simeq
\left\langle
\Delta u
\right\rangle_{B,\alpha,T_A}
-
\left\langle
\Delta u
\right\rangle_{B,\beta,T_A},
\label{eq:matchgeneral}
\end{equation}

where $\Delta u=\Delta U/N$ and the averages are taken in the isothermal isobaric ensemble of the base potential.
Equation~\eqref{eq:matchgeneral} is independent of the functional form of $\Delta U$.
Read as a residual estimate, it would require a potential for description $A$, which is not available for an assessed reference.
We use it in reverse, with $\Delta U$ as the unknown.
The correction potential is fitted so that its phase averaged energy difference reproduces the residual.
The derivations, error terms, and relative merits of the residual estimates are discussed in Sec.~\ref{sec:fediff}.

In this work, we use Eq.~\eqref{eq:targetB} to construct the thermodynamic residual and Eq.~\eqref{eq:matchgeneral} to translate this residual into an energy target for the additive correction layer $\Delta U$.
In practice, the averages in Eq.~\eqref{eq:matchgeneral} are taken at the measured transformation temperature $T_k$ of the current potential (Sec.~\ref{sec:fitting}) and at the neighboring sampled temperatures, with the same target at each temperature.
Because the correction also changes the transition entropy, the shift obtained at $T_A$ differs from the target by $-\delta s_{\alpha\beta}(T_A-T_k)$, where $\delta s_{\alpha\beta}$ is the change of the transition entropy produced by the correction at a common temperature.
Since $\delta s_{\alpha\beta}$ is proportional to the correction, and the correction to $T_A-T_k$, this term is second order in $T_A-T_k$.
It is reduced in the next correction round, which is sampled at the new transformation temperature, together with the heat capacity term of Eq.~\eqref{eq:gexpand} and the second cumulant of Eq.~\eqref{eq:secondcumulant}.
After each correction step, the corrected phase free energies are measured again.
If the transformation temperature remains outside the prescribed tolerance, Eq.~\eqref{eq:targetB} is applied to the current corrected potential to construct the next residual.
The correction has no explicit temperature dependence.
Its thermodynamic contribution changes with temperature through the distribution of atomic environments sampled by each phase.
The reference transformation temperatures are taken from the SGTE description of the pure elements \cite{Dinsdale1991}.
The same procedure can be applied using another thermodynamic assessment or directly measured transformation temperatures.


\subsection{Thermodynamic response and iterative correction}
\label{sec:fitting}

\begin{figure*}[htp!]
    \centering
    \includegraphics[width=0.8\linewidth]{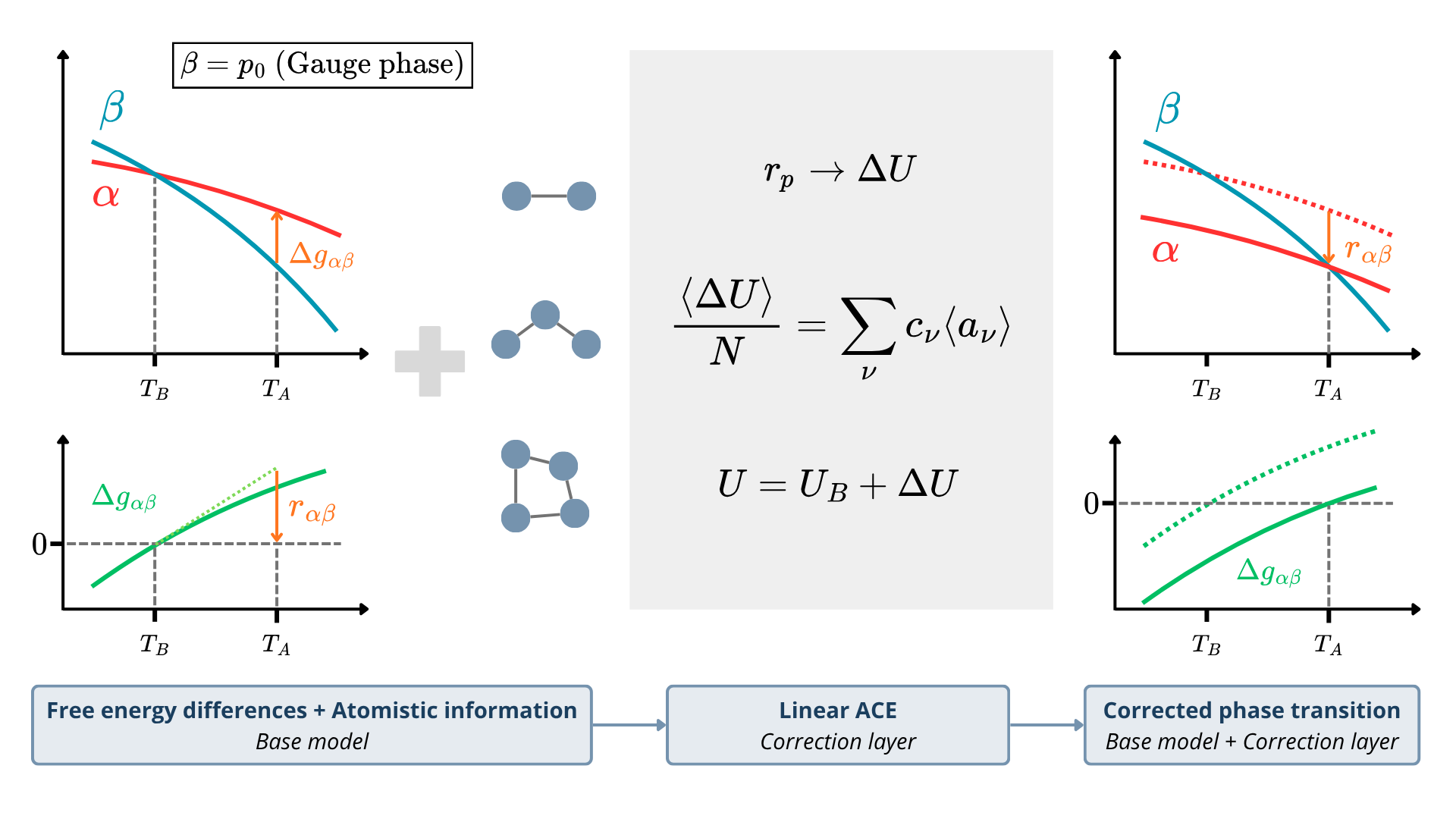}
    \caption{
    Workflow for shifting the transformation temperature between phases $\alpha$ and $\beta$ from the value predicted by the base potential, $T_B$, towards the reference value, $T_A$.
    The correction layer is fitted iteratively from measured phase free energies, as described in Sec.~\ref{sec:fitting}.
    }
    \label{fig:workflow}
\end{figure*}

Irrespective of the functional form of $U_B$, we represent $\Delta U$ by a linear atomic cluster expansion (ACE) containing the same chemical species as the base potential \cite{Drautz2019},

\begin{equation}
\Delta U(\mathbf{R})
=
\sum_{i=1}^{N}
\sum_{\nu}
c_\nu B_\nu(\sigma_i),
\label{eq:ansatz}
\end{equation}

where $\mathbf{R}$ denotes an atomic configuration, $\sigma_i$ is the local environment of atom $i$, $B_\nu$ are the ACE basis functions, and $c_\nu$ are the expansion coefficients.

For each configuration, we define the per atom basis sums
$a_\nu(\mathbf{R}) = N^{-1} \sum_{i=1}^{N} B_\nu(\sigma_i)$,
such that

\begin{equation}
\left\langle
\Delta u
\right\rangle_{B,p,T}
=
\sum_\nu
c_\nu
\left\langle
a_\nu
\right\rangle_{B,p,T}.
\label{eq:meanresponse}
\end{equation}

Inserting Eq.~\eqref{eq:meanresponse} into Eq.~\eqref{eq:matchgeneral}, with the averages taken at the sampled temperature $T$, gives

\begin{equation}
r_{\alpha\beta}
\simeq
\sum_\nu
c_\nu
\left[
\left\langle
a_\nu
\right\rangle_{B,\alpha,T}
-
\left\langle
a_\nu
\right\rangle_{B,\beta,T}
\right].
\label{eq:linearresponse}
\end{equation}

Equation~\eqref{eq:linearresponse} is linear in the ACE coefficients.
We solve this relation using a regularized constrained least squares fit that limits the magnitude and forces of the correction and imposes stability bounds outside the thermodynamic fitting configurations.
The complete fitting procedure is described in Sec.~\ref{sec:fitmethods}.

A single correction does not necessarily place the transformation temperature exactly at the reference value.
The thermodynamic residual is a local first order estimate, while the free energy response of the correction is also approximated to first order and is subject to the regularization and stability constraints of the fit.
We therefore refine the correction using measured phase free energies.

After $k$ correction steps, the combined potential is $U_k=U_B+\Delta U_k$,
where $\Delta U_k$ is the accumulated correction.
The phase free energies of $U_k$ are calculated as described in Sec.~\ref{sec:rs}, giving a measured transformation temperature $T_k$ and transition entropy $\Delta s_{\alpha\beta}^{(k)}$.
If $T_k$ differs from the reference value $T_A$, the remaining free energy shift $r_{\alpha\beta,k}$ is estimated from Eq.~\eqref{eq:targetB} with $T_B$ and $\Delta s_{\alpha\beta}^{B}$ replaced by $T_k$ and $\Delta s_{\alpha\beta}^{(k)}$.
A new correction increment $\delta U_k$ is then fitted by applying Eq.~\eqref{eq:linearresponse} to configurations sampled with the current potential $U_k$, with the coefficients $c_\nu$ replaced by the increments $\delta c_{k,\nu}$ and the target by $r_{\alpha\beta,k}$.
The accumulated correction is updated as $\Delta U_{k+1}=\Delta U_k+\delta U_k$,
and the phase free energies are then measured again.
The correction procedure is repeated until all transformation temperatures of the element lie within a prescribed tolerance of $T_A$.

The iteration can be viewed as a local Newton update on the transformation temperature, with the measured transition entropy providing the local thermodynamic slope.
Convergence requires that the same two phases remain connected by the same transformation over the temperature interval considered and that the transition entropy remains finite.
Because every round uses the measured transformation temperature and transition entropy of the current potential, errors in the first order perturbative estimate or in the fitted realization of the requested shift affect the number of rounds rather than the final convergence criterion.

Details of the free energy measurements, sampling procedure, and numerical fitting are given in Secs.~\ref{sec:rs}--\ref{sec:fitmethods}. The methodology used to construct the iterative correction layer described above is summarized in Fig.~\ref{fig:workflow}.


\subsection{Application to a foundation MLIP}
\label{sec:application}

We demonstrate the correction procedure using the GRACE foundation model, specifically the \texttt{GRACE-2L-SMAX-OMAT-large} model \cite{Bochkarev2026} based on the graph atomic cluster expansion architecture \cite{Bochkarev2024,Lysogorskiy2026Foundation}.
The test set contains Ag, Al, Au, Ca, Cu, Mg, Mo, Ni, Ti, V, and Zr.
It comprises melting transitions together with the fcc to bcc transition of Ca and the hcp to bcc transitions of Ti and Zr.

\begin{figure*}[htp!]
    \centering
    \includegraphics[width=1.0\linewidth]{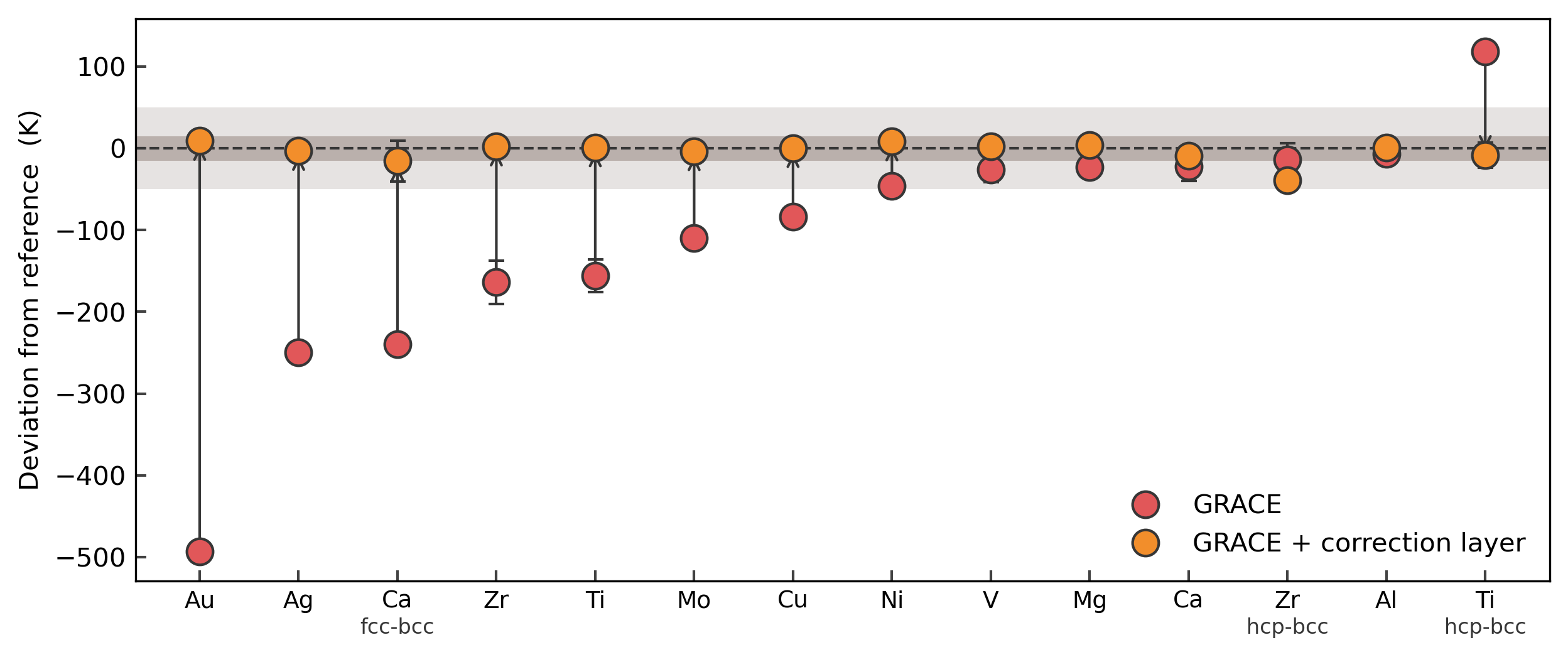}
    \caption{
    Deviation from the SGTE transformation temperatures for the base GRACE model with and without the converged correction layer.
    Arrows connect the uncorrected and corrected values.
    Shaded regions indicate the tolerances of 15~K for melting and 50~K for solid to solid transitions.
    Error bars are standard deviations over independent sweeps (Table~\ref{tab:results}).
    }
    \label{fig:results}
\end{figure*}

Each element is treated independently.
The phase free energies and transformation temperatures of the unmodified GRACE potential are calculated using the reversible scaling procedure described in Sec.~\ref{sec:rs}.
Figure~\ref{fig:results} and Table~\ref{tab:results} compare the resulting transformation temperatures with the SGTE reference values.
The transformation temperatures predicted by GRACE show a broad range of deviations from the SGTE values.
All melting temperatures of GRACE are lower than the assessed values, while several transitions are already reproduced closely.
The \texttt{GRACE-2L-SMAX-OMAT-large} model is trained on PBE reference data from SMAX \cite{Bochkarev2026} as well as from OMat24 \cite{BarrosLuque2026}.

\begin{table*}
    \caption{
Transition temperatures and transition entropies for the assessed reference, the base model, and the corrected model.
SGTE denotes the assessed values from the SGTE unary database, the reference description $A$ with $T_A$ and $\Delta S_A$.
GRACE denotes values measured for the \texttt{GRACE-2L-SMAX-OMAT-large} base model, the description $B$ with $T_B$ and $\Delta S_B$.
$\Delta S=N_{\mathrm{A}}(s_\alpha-s_\beta)=N_{\mathrm{A}}\Delta s_{\alpha\beta}$ is the molar transition entropy, negative for every transition.
Corrected denotes values measured after adding the correction layer.
The GRACE values are the mean $\pm$ standard deviation over five independent reversible scaling sweeps, and the corrected values over three independent sweeps of the final layer.
$N$ is the number of correction rounds, each followed by a free energy measurement.
    }
\label{tab:results}
    \centering
\small
\begin{tabular}{llrrrcrrr}
\toprule
 & & \multicolumn{3}{c}{$T$ (K)} & & \multicolumn{3}{c}{$\Delta S$ (J\,mol$^{-1}$\,K$^{-1}$)} \\
\cmidrule(lr){3-5}\cmidrule(lr){7-9}
Element & Transition & SGTE & GRACE & corrected & $N$ & SGTE & GRACE & corrected \\
\midrule
Al & melting & 933 & 927 $\pm$ 2 & 934 $\pm$ 1 & 1 & $-11.48$ & $-11.27 \pm 0.07$ & $-11.27 \pm 0.04$ \\
Cu & melting & 1358 & 1274 $\pm$ 2 & 1358 $\pm$ 1 & 1 & $-9.77$ & $-9.45 \pm 0.09$ & $-9.55 \pm 0.02$ \\
Ni & melting & 1728 & 1682 $\pm$ 2 & 1737 $\pm$ 5 & 1 & $-10.11$ & $-10.47 \pm 0.04$ & $-10.41 \pm 0.04$ \\
Ag & melting & 1235 & 985 $\pm$ 7 & 1232 $\pm$ 3 & 1 & $-9.15$ & $-9.48 \pm 0.04$ & $-8.63 \pm 0.04$ \\
Au & melting & 1337 & 844 $\pm$ 4 & 1346 $\pm$ 2 & 2 & $-9.39$ & $-9.34 \pm 0.07$ & $-8.30 \pm 0.04$ \\
Mg & melting & 923 & 900 $\pm$ 4 & 927 $\pm$ 3 & 1 & $-9.18$ & $-9.30 \pm 0.04$ & $-9.19 \pm 0.04$ \\
\addlinespace
\multirow{2}{*}{Ca} & fcc$\to$bcc & 716 & 477 $\pm$ 12 & 700 $\pm$ 25 & 3 & $-1.30$ & $-1.01 \pm 0.05$ & $-1.11 \pm 0.04$ \\
 & melting & 1115 & 1093 $\pm$ 18 & 1106 $\pm$ 5 & 3 & $-7.66$ & $-7.3 \pm 0.2$ & $-7.26 \pm 0.02$ \\
\addlinespace
\multirow{2}{*}{Ti} & hcp$\to$bcc & 1155 & 1273 $\pm$ 9 & 1147 $\pm$ 16 & 2 & $-3.61$ & $-2.83 \pm 0.07$ & $-3.19 \pm 0.04$ \\
 & melting & 1941 & 1785 $\pm$ 20 & 1942 $\pm$ 10 & 2 & $-7.28$ & $-6.37 \pm 0.09$ & $-6.78 \pm 0.01$ \\
\addlinespace
\multirow{2}{*}{Zr} & hcp$\to$bcc & 1139 & 1126 $\pm$ 20 & 1100 $\pm$ 9 & 1 & $-3.61$ & $-2.9 \pm 0.1$ & $-3.23 \pm 0.07$ \\
 & melting & 2128 & 1964 $\pm$ 27 & 2130 $\pm$ 5 & 1 & $-9.87$ & $-6.5 \pm 0.2$ & $-6.58 \pm 0.02$ \\
\addlinespace
V & melting & 2183 & 2157 $\pm$ 16 & 2186 $\pm$ 3 & 1 & $-9.85$ & $-10.5 \pm 0.1$ & $-10.61 \pm 0.03$ \\
Mo & melting & 2896 & 2786 $\pm$ 11 & 2892 $\pm$ 5 & 1 & $-12.94$ & $-13.3 \pm 0.1$ & $-13.00 \pm 0.05$ \\
\bottomrule
\end{tabular}
\end{table*}

The largest uncorrected deviation occurs for Au.
This is consistent with our previous Au--Cu phase diagram calculations, where phase equilibria were sensitive to the electronic structure reference, with different results obtained using LDA, PBE, and r2SCAN \cite{Menon2026}.
Independent studies show that GGA gives a poor thermodynamic description of Au at finite temperature, with excessive thermal softening and an instability below the experimental melting point \cite{Grabowski2015RPA}.
More generally, finite temperature free energies of fcc metals are sensitive to exchange correlation effects and anharmonic contributions \cite{Grabowski2007,Glensk2015}.
The Au case therefore provides a stringent test for the correction layer, since the required shift is large.

The corrected values are shown by the orange symbols in Fig.~\ref{fig:results} and listed in Table~\ref{tab:results}.
All transitions are brought within the prescribed tolerance, including systems for which the initial difference from the assessment is several hundred kelvin.
The Zr hcp to bcc transition is the only transition that moves away from its reference.
It is within the tolerance for the base model and moves further from the assessed value when the melting transition is corrected with the same layer.
It remains within the tolerance for solid to solid transitions.
A single correction layer can also treat several phases of the same element.
For Ca, Ti, and Zr, the solid to solid and melting transitions are corrected simultaneously using one set of ACE coefficients.
Except for the Zr hcp to bcc transition discussed above, the remaining differences from the SGTE values are comparable to the scatter between independent reversible scaling sweeps.
The reported error bars are obtained from independent reversible scaling sweeps of the final corrected layer.

\begin{figure*}[htp!]
    \centering
    \includegraphics[width=1.0\linewidth]{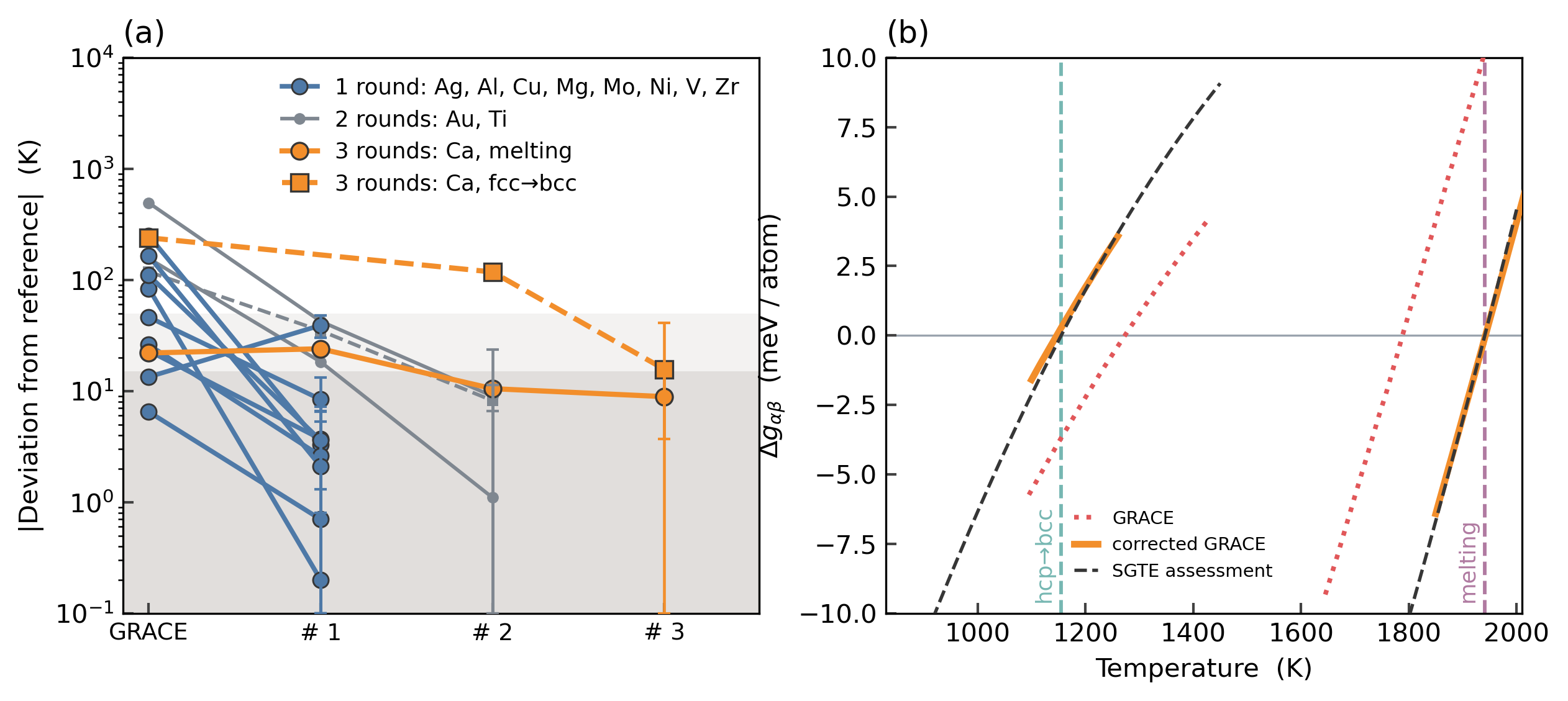}
    \caption{
    Convergence of the thermodynamic correction.
    (a) Absolute deviation from the reference transformation temperature as a function of correction round.
    Grey lines show all transitions, blue lines those that converge in one round, and orange lines those that need the most rounds.
    Shaded regions indicate the convergence tolerances.
    The last point of each transition is the mean over three sweeps of the final layer.
    (b) Ti free energy differences $\Delta g_{\alpha\beta}$, low temperature phase minus high temperature phase, for GRACE, corrected GRACE, and the SGTE assessment.
    Dashed lines indicate the SGTE transformation temperatures.
    }
    \label{fig:fevsiter}
\end{figure*}

Figure~\ref{fig:fevsiter}(a) shows the measured convergence of the transformation temperatures.
Several elements enter the melting tolerance after a single correction round, while systems requiring larger free energy shifts need additional iterations.
The shallow solid to solid transitions are more sensitive to small free energy changes and can also require several rounds.

Figure~\ref{fig:fevsiter}(b) illustrates how the correction modifies the free energy branches for Ti.
GRACE provides the free energy branches for the hcp, bcc, and liquid phases, while the correction shifts their relative levels so that both transformation temperatures move towards the SGTE values.
For the bcc to liquid transition, the corrected free energy difference follows the SGTE assessment closely over the sampled temperature range even though the assessed entropy is not fitted explicitly.
For the hcp to bcc transition, the crossing is reproduced and the slope moves from the GRACE value towards the SGTE value, but remains below it.

\begin{figure*}[htp!]
    \centering
    \includegraphics[width=1.0\linewidth]{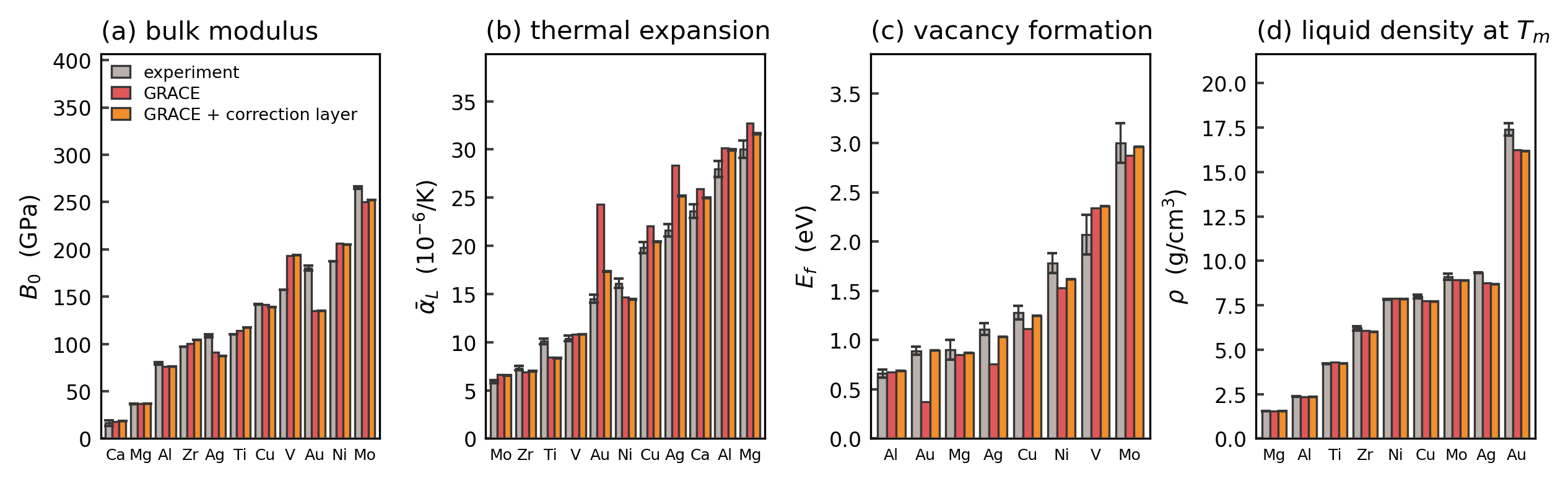}
\caption{Properties of \texttt{GRACE-2L-SMAX-OMAT-large} with and without the correction layer, compared with experiment.
(a) Bulk modulus~\cite{Neighbours1958,Kamm1964,Overton1955,Slutsky1957,Featherston1963,Alers1960Ni,Fisher1964,Alers1960V,Anzellini2018}.
(b) Mean linear thermal expansion coefficient from 293~K~\cite{Touloukian1975}.
(c) Vacancy formation energy~\cite{Schaefer1987}.
(d) Liquid density at the melting temperature of each potential~\cite{Abdullaev2020,Paradis2008,Assael2006,Assael2010,Abdullaev2015,Abdullaev2019,Paradis2002,Paradis2000,Ishikawa2005}.
Elements are ordered by the experimental value.}
    \label{fig:properties}
\end{figure*}

We next examine whether the correction degrades properties of the underlying GRACE model.
All elements are evaluated using the procedures described in Sec.~\ref{sec:validation}.
Figure~\ref{fig:properties} compares the bulk modulus, linear thermal expansion coefficient, vacancy formation energy, and liquid density at the melting temperature.
Only elements with available reference data are included in this comparison.

The bulk moduli change only weakly after adding the correction layer.
The base model gives values within 10\% of experiment for most elements.
For V, both potentials give a negative $C_{44}$ for bcc at 0~K, so its bulk modulus refers to a mechanically unstable structure.
In addition, the full energy volume curves of the relevant polymorphs are shown in Fig.~S1.
They retain smooth minima and the same zero temperature phase ordering as the base model.
The linear thermal expansion coefficients change little for most elements.
The exceptions are Au, Ag, and Cu, where the correction lowers the coefficient towards the experimental value.
The underlying volume curves used to obtain the thermal expansion coefficients are shown in Fig.~S2.

Vacancy formation energies increase with the correction, most clearly for Au, Ag, Cu, Ni, and Mo.
The correction raises the energy of the liquid relative to the crystal, and the under-coordinated atoms next to a vacancy are raised with it, which could be a reason for this change.
Empirical correlations between vacancy formation energy and melting temperature have been reported for metals \cite{Hayashiuchi1982,Grimvall1999,Kraftmakher1998}.
Since GRACE underestimates the vacancy formation energies of these elements, the correction layer moves them towards experiment.
For V and Al, GRACE already lies above experiment, and the small increase moves them away from it.
The liquid densities at the melting temperature are also nearly unchanged by the correction.
For Au and Ag, the melting temperatures of the two potentials differ by about 500 and 250~K, and the thermal expansion of the liquid over this interval offsets the effect of the correction.
At a common temperature of $1.03\,T_A$ (Table~S2), the correction increases the liquid density of Au and Ag by about 3 to 6\%, towards experiment, and changes it by 1\% or less for the other elements.
The radial distribution functions of the liquids, shown in Fig.~S3, confirm that the local liquid structure remains close to that of the base model.
These tests show that the correction layer can modify phase thermodynamics without replacing the atomistic description provided by the foundation model.
The effect on independent properties depends on how the correction couples to the sampled local environments, and is not determined only by the size of the transition temperature shift.

\begin{figure*}[htp!]
    \centering
    \includegraphics[width=0.8\linewidth]{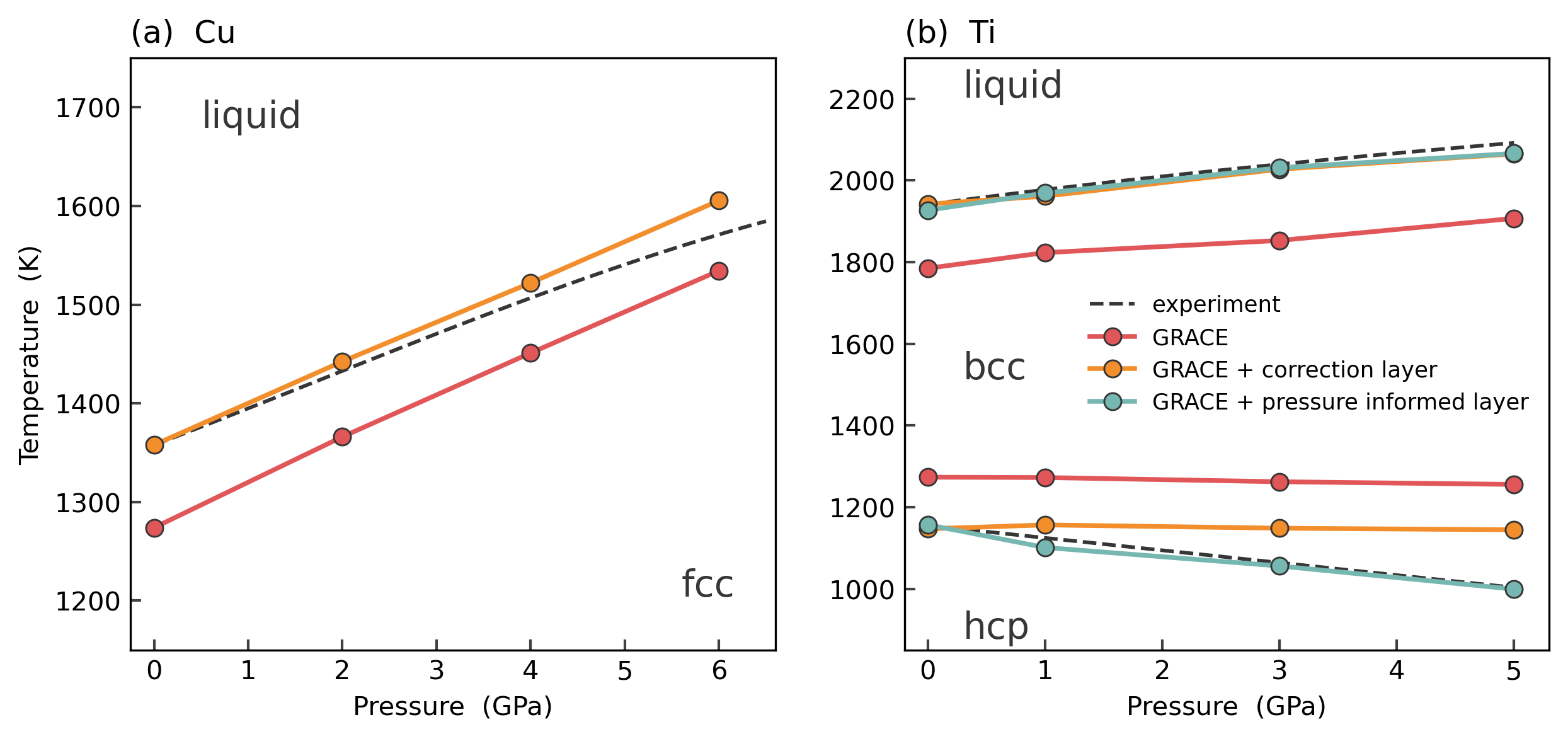}
    \caption{
    Pressure temperature phase boundaries of (a) Cu and (b) Ti.
    Symbols show GRACE, GRACE with the correction layer fitted at zero pressure, and, for Ti, GRACE with a pressure informed correction layer.
    Dashed lines show the fit to the measured melting temperatures of Cu~\cite{Akella1971}, the Simon fit to the measured melting temperatures of Ti~\cite{Stutzmann2015}, and the hcp to bcc boundary of Ti~\cite{Tonkov2004} as compiled in Ref.~\cite{Wen2021}.
    The $\omega$ phase of Ti is not considered.}
    \label{fig:ptphase}
\end{figure*}

We finally examine the extension of the correction to finite pressure.
Figure~\ref{fig:ptphase} shows pressure temperature phase boundaries for Cu and Ti.
At each pressure, the transformation temperatures are obtained from reversible scaling calculations using the same free energy procedure as at zero pressure.

For Cu, the correction fitted only at zero pressure follows the melting curve of Akella and Kennedy within 15~K up to 4~GPa and lies 35~K above it at 6~GPa~\cite{Akella1971}.
Earlier measurements lie up to about 60~K higher at these pressures~\cite{Akella1971}, so the deviation is within the spread of the experimental data.
The correction shifts the absolute melting temperature while retaining the pressure dependence represented by GRACE.

Ti provides a complementary example.
The zero pressure correction moves both transformation temperatures towards experiment at ambient pressure and retains the pressure dependence of the underlying GRACE model.
For melting, this gives good agreement over the pressure interval considered.
For the hcp to bcc transition, the pressure dependence of the GRACE model differs from the experimental boundary, so correcting the zero pressure crossing alone does not determine the complete phase line.

The correction can incorporate this additional thermodynamic information directly.
We therefore refit the Ti layer using transformation temperatures at several pressures together with the zero pressure targets.
The resulting pressure informed layer reproduces both the melting and hcp to bcc boundaries over the sampled pressure range.
This demonstrates that the correction can retain thermodynamic trends already captured by the foundation model, as for Cu, while incorporating additional pressure dependent information when required.
The same correction layer can therefore represent multiple solid and liquid phases together with transformation data over both temperature and pressure.


\section{Discussion}

The thermodynamic correction layer provides a way to combine a transferable interatomic potential with selected thermodynamic reference data without modifying the underlying model.
The base potential supplies the atomistic description and the temperature dependence of the phase free energies, while the correction adjusts their relative levels to reproduce target transformation temperatures.
Because phase coexistence is determined by small free energy differences, comparatively small energetic corrections can produce substantial changes in transformation temperature.

The residual prescribes the free energy shift required to move the phase crossing.
The assessed transition entropy is not imposed in the fit.
Applying the same target over several sampled temperatures favors an approximately rigid shift of the relative phase free energies.
The transition entropies reported in Table~\ref{tab:results} therefore provide an independent check of the thermodynamic response of the corrected model.
For several transitions, the corrected entropy remains close to the base value.
For others, most notably Ag and Au melting, the entropy changes more visibly.
For the solid to solid transitions of Ca, Ti, and Zr and for Ti melting, the correction moves the transition entropy towards the assessed value.
For Ag and Au, the correction is large enough to change the configurations sampled in each phase.
This change is beyond the first order fit and also appears in the liquid density at a common temperature.
Where the base transition entropy differs strongly from the assessed one, the corrected crossing is correct but the free energy difference deviates from the assessment away from $T_A$.
For Zr melting, the corrected transition entropy remains about one third smaller in magnitude than the assessed value.

Because convergence is judged from newly measured free energies, the same procedure treats transitions with substantially different free energy slopes, including melting and solid to solid transformations.
The correction retains the static energetics and liquid structure of the underlying potential, consistent with the intended role of the layer as a perturbation of the phase balance rather than a replacement of the atomistic model.
Properties linked to the phase balance, such as vacancy formation energies and thermal expansion, change where the correction is large and mostly towards experiment.
It can preserve thermodynamic trends already present in the base model while incorporating additional pressure dependent information where required, as the Cu and Ti phase boundaries show.

An important feature of the method is that the base potential itself does not require retraining or modification.
Only energies and configurations generated with the potential are needed to construct the correction.
The architecture, parameters, and original training data of the model do not need to be accessible.
The procedure can therefore in principle be applied to universal machine learning interatomic potentials, specialized machine learning potentials, and conventional interatomic potentials.
This is particularly useful for universal models, where retraining the complete potential for a specific thermodynamic application would be substantially more expensive than fitting a small additive layer.
The cost of the procedure is dominated by the free energy calculations.
For Ti, the two correction rounds, including sampling and free energy calculations, required about 55 GPU hours on NVIDIA H100 GPUs, about 80\% of which were free energy calculations.
The fits run on a CPU node and add negligible cost.

There are limits to what can be represented by a temperature independent correction of the atomic energy.
The layer can rebalance phases that are already represented by the base potential, but it cannot introduce physical degrees of freedom that are absent from that description.
Magnetic phase stability is a clear example.
If the relevant thermodynamic driving force contains magnetic entropy or magnetic order that is absent from the base model, an atomic energy correction can only compensate the projected effect within the sampled configurations.
It cannot create the missing magnetic degrees of freedom.
Similar limitations can arise when a phase transformation depends strongly on electronic or other degrees of freedom that are not contained in the underlying model.

The corrected transformation temperatures are calibrated quantities rather than independent predictions because thermodynamic reference data enter the fitting targets.
Their accuracy is consequently limited by the quality of the reference data and by the statistical and finite size uncertainties of the free energy calculations.
The correction is also constrained only over the configuration space represented by the thermodynamic sampling and stability calculations.
Applications outside this region require the same validation that would be expected for the underlying interatomic potential.
Where the energies of specific polymorphs matter, their static configurations can be added to the fit as constraints, as done here for the crystal structures of Ag and Au.
The stability constraints, the random structures, dimers, and harvested environments, enter the fit only as bounds on the correction energy and require no energy targets from the base model or from a reference electronic structure method.
Their generation and screening are therefore inexpensive compared with the free energy calculations.
These constraints are nevertheless required as the thermodynamic targets fix only phase averaged energy differences and leave most directions of the coefficient space undetermined.

The method requires neither differentiable molecular dynamics nor gradients through the free energy calculation.
It also does not require access to the trainable parameters of the base model.
It is therefore complementary to approaches based on differentiable simulations or direct optimization of thermodynamic observables \cite{Thaler2021,Fuchs2026,Jin2026}, to the direct incorporation of free energies in parameter fitting \cite{Wei2026}, and to additive corrections on fixed potentials trained on structural data \cite{Matin2024}.
The closest recent examples are differentiable refinement of Ti phase boundaries by Fuchs and Zavadlav \cite{Fuchs2026} and descriptor based refinement of an Fe transformation temperature by Swinburne, Lapointe, and Marinica \cite{Swinburne2025}.
For Ti, Fuchs and Zavadlav obtain both boundaries within 50~K up to 5~GPa by refining the parameters of the potential, which increases its energy error with respect to the DFT reference \cite{Fuchs2026}.
The pressure informed correction layer reaches a comparable accuracy while leaving the base potential itself unchanged.
Those works show that phase boundaries can be targeted directly through free energy information.
The present work uses the same thermodynamic principle in a different setting.
The base potential is treated as a fixed interaction model, and the thermodynamic information is converted into energy labels for an additive linear correction layer.
The construction is therefore independent of the architecture and trainable parameters of the base model.

The present work focuses on elemental phase transformations, but the formulation is not restricted to unary systems.
For alloys, composition dependent phase free energy differences could be incorporated using a multicomponent correction basis.
Together with the finite pressure extension demonstrated here, this provides a route towards corrections defined over composition, temperature, and pressure.
Such layers could connect universal atomistic models with assessed phase thermodynamics while retaining the atomistic resolution and configurational coverage of the underlying potential.
Electronic structure calculations provide a microscopic energy model, but phase stability remains sensitive to the chosen density functional.
Experimental assessments define the thermodynamic reference, but they do not provide an atomistic potential.
A thermodynamic correction layer separates these roles by preserving the microscopic representation of the base potential while fitting its phase thermodynamics to assessed data.

\section{Methods}
\label{sec:methods}


\subsection{Free energy differences}
\label{sec:fediff}

This section derives the residual estimates introduced in Sec.~\ref{sec:correction} and relates them to the energy targets used to fit the correction layer.

For each description, the free energy difference between the phases is $\Delta g_{\alpha\beta}(T)=\Delta u_{\alpha\beta}(T)-T\Delta s_{\alpha\beta}(T)$,
and the crossing condition $\Delta g_{\alpha\beta}^{M}(T_M)=0$ gives

\begin{equation}
\Delta s_{\alpha\beta}^{M}(T_M)
=
\frac{
\Delta u_{\alpha\beta}^{M}(T_M)
}{
T_M
},
\qquad
M=A,B.
\label{eq:crossrel}
\end{equation}

Around its own crossing, the free energy difference of description $M$ can be expanded as

\begin{equation}
\begin{aligned}
\Delta g_{\alpha\beta}^{M}(T)
&=
-
\Delta s_{\alpha\beta}^{M}(T_M)
\left(
T
-
T_M
\right)
\\
&\quad
-
\frac{
\Delta c_{\alpha\beta}^{M}(T_M)
}{
2T_M
    }
\left(
T
-
T_M
\right)^2
+
\ldots,
\end{aligned}
\label{eq:gexpand}
\end{equation}

where we used $\mathrm{d}\Delta g/\mathrm{d}T=-\Delta s$ and $\mathrm{d}^2\Delta g/\mathrm{d}T^2=-\Delta c/T$.
Here, $\Delta c_{\alpha\beta}$ is the heat capacity difference between the phases.

Evaluating Eq.~\eqref{eq:gexpand} for the base potential at $T_A$ and using Eq.~\eqref{eq:crossrel} gives Eq.~\eqref{eq:targetB}.
The leading omitted term in the residual is $\Delta c_{\alpha\beta}^{B}(T_B)(T_A-T_B)^2/(2T_B)$.

This construction does not assume matching entropies between the two descriptions.
Through the crossing relation, the plain ensemble average $\Delta u_{\alpha\beta}^{B}(T_B)$ replaces the numerical derivative of the free energy branches.
Mixed evaluations, such as using $\Delta s^{B}$ at $T_B$ together with $\Delta u^{B}$ at $T_A$ in the expression for $\Delta g_{\alpha\beta}$, leave errors that are first order in $T_A-T_B$.

For the construction from energy differences, evaluating $\Delta g_{\alpha\beta}=\Delta u_{\alpha\beta}-T\Delta s_{\alpha\beta}$ at $T_A$, adding and subtracting $T_A\Delta s_{\alpha\beta}^{A}(T_A)$, and using the crossing relation of description $A$ gives

\begin{equation}
\begin{aligned}
r_{\alpha\beta}
&=
\Delta u_{\alpha\beta}^{A}(T_A)
-
\Delta u_{\alpha\beta}^{B}(T_A)
\\
&\quad
-
T_A
\left[
\Delta s_{\alpha\beta}^{A}(T_A)
-
\Delta s_{\alpha\beta}^{B}(T_A)
\right].
\end{aligned}
\label{eq:targetAfull}
\end{equation}

Dropping the entropy mismatch term gives

\begin{equation}
r_{\alpha\beta}
\simeq
\Delta u_{\alpha\beta}^{A}(T_A)
-
\Delta u_{\alpha\beta}^{B}(T_A).
\label{eq:targetA}
\end{equation}

This approximation compares the transition entropies of both descriptions at the same temperature $T_A$.
The omitted term is linear in the entropy mismatch and carries the prefactor $T_A$, so it can be comparable to the residual where the entropies differ.

The first order form of Eq.~\eqref{eq:targetB} is the relation used by Alf\`e, Gillan, and Price to transfer melting temperatures between two energy models \cite{Alfe2002}.
Here, we use the same relation in the opposite direction, from a prescribed temperature shift to the required free energy shift.
Sturgeon and Laird earlier adjusted an embedded atom potential to the experimental melting point of aluminium using a related thermodynamic argument \cite{Sturgeon2000}.

The three residual constructions differ in the information they require.
The direct evaluation in Eq.~\eqref{eq:targetdirect} makes no approximation but requires measured free energy branches at $T_A$.
Equation~\eqref{eq:targetB} requires a free energy calculation to locate $T_B$, but from the reference it requires only the transformation temperature $T_A$.
It therefore applies whenever an experiment provides a single measured temperature, such as one point on a pressure temperature phase boundary, without direct access to $u^A$, $s^A$, or $g^A$.
Equation~\eqref{eq:targetA} requires the assessed transition enthalpy but no free energy calculation for the base potential, and is the cheapest construction when calorimetric data or a thermodynamic assessment are available.
Errors of the approximate constructions are absorbed by the iterative refinement.

The first order response of the free energy to the correction follows from a variational argument.
At fixed temperature and pressure, the Gibbs free energy of a phase is a functional of the normalized configurational distribution $\rho$,

\begin{equation}
G[\rho]
=
\left\langle
U
+
pV
\right\rangle_{\rho}
+
k_{\mathrm{B}}T
\left\langle
\ln \rho
\right\rangle_{\rho},
\label{eq:gfunctional}
\end{equation}

up to terms independent of the potential energy.
The functional is stationary at the equilibrium distribution of the base potential,
$\rho_B \propto \exp[-(U_B+pV)/(k_{\mathrm{B}}T)]$.
For the corrected potential, $G^C[\rho]=G^B[\rho]+\left\langle \Delta U \right\rangle_{\rho}$.
This follows because, at a fixed distribution $\rho$, the potential energy enters Eq.~\eqref{eq:gfunctional} only through the average $\langle U\rangle_{\rho}$, and replacing $U_B$ by $U_C=U_B+\Delta U$ adds $\langle \Delta U\rangle_{\rho}$.
The pressure term $p\langle V\rangle_{\rho}$ and the entropy term $k_{\mathrm{B}}T\langle \ln\rho\rangle_{\rho}$ depend only on $\rho$, $p$ and $T$, which are the same for both functionals, and therefore cancel in the difference $G^C[\rho]-G^B[\rho]$.
Changes of volume and entropy enter only through the change of the distribution from $\rho_B$ to $\rho_C$, which is treated next.
Evaluating this expression at the perturbed equilibrium distribution $\rho_C=\rho_B+\delta\rho$, with $\delta\rho$ first order in $\Delta U$, the stationarity of $G^B$ removes the term linear in $\delta\rho$.
For each phase,

\begin{equation}
G^C
=
G^B
+
\left\langle
\Delta U
\right\rangle_B
+
\mathcal{O}\!\left(
\Delta U^2
\right).
\label{eq:stationary}
\end{equation}

The distribution, and therefore the entropy and volume of each phase, change at first order in $\Delta U$.
The free energy is insensitive to this change at first order because it is stationary with respect to the distribution.
Dividing Eq.~\eqref{eq:stationary} by the number of atoms of each phase and taking the phase difference gives Eq.~\eqref{eq:matchgeneral}.
The leading neglected term is the second cumulant,

\begin{equation}
-
\frac{
\left\langle
\Delta U^2
\right\rangle_B
-
\left\langle
\Delta U
\right\rangle_B^2
}{
2 k_{\mathrm{B}}T
},
\label{eq:secondcumulant}
\end{equation}

with higher order terms following from the same expansion.
Per atom, as in Eq.~\eqref{eq:matchgeneral}, this term is divided by the number of atoms $N$ of the phase.
These contributions are not fitted explicitly in the present work, but are absorbed by the iterative refinement.
The perturbative evaluation of free energy shifts from configuration energy differences was used in the same context by Alf\`e, Gillan, and Price \cite{Alfe2002}.

Although Eq.~\eqref{eq:matchgeneral} is written in terms of the average correction energy, its first order meaning is the change in the phase free energy difference.
The corresponding change in the phase internal energy difference also contains the entropy response, as in Eq.~\eqref{eq:targetAfull} with $A$ replaced by $C$ and $r_{\alpha\beta}$ replaced by the first order change $\Delta\Delta g_{\alpha\beta}(T)$ of the phase free energy difference generated by the correction.

To first order, the associated entropy change is $\Delta s_{\alpha\beta}^{C}-\Delta s_{\alpha\beta}^{B}=-\mathrm{d}\Delta\Delta g_{\alpha\beta}/\mathrm{d}T$.
It vanishes when the correction produces a temperature independent shift of the phase free energy difference.
Applying the same target at several sampled temperatures therefore favors a rigid shift of the phase free energy difference and limits changes to the transition entropy within the first order fit.


\subsection{Free energy calculations and reversible scaling}
\label{sec:rs}

Phase free energies are calculated using the nonequilibrium methods implemented in \texttt{calphy} \cite{Menon2021}.
For each phase, an absolute free energy is first determined at a reference temperature $T_0$ by nonequilibrium Hamiltonian interpolation \cite{Freitas2016}.
An Einstein crystal is used as the reference system for solids \cite{FrenkelLadd1984}, while the Uhlenbeck and Ford model is used for liquids \cite{PaulaLeite2016}.

The system is switched continuously between the interatomic potential and the corresponding reference system in both directions.
Using the work per atom $w_{\mathrm{f}}$ and $w_{\mathrm{b}}$ from the forward and backward transformations, respectively, the free energy per atom is estimated as

\begin{equation}
g(T_0)
=
g_{\mathrm{ref}}(T_0)
+
\frac{1}{2}
\left[
w_{\mathrm{f}}
-
w_{\mathrm{b}}
\right],
\label{eq:fl}
\end{equation}

with the corresponding dissipation $e_{\mathrm{diss}}=\frac{1}{2}[w_{\mathrm{f}}+w_{\mathrm{b}}]\ge 0$.
For solids, the Einstein crystal reference includes the center of mass correction.
At finite pressure, the absolute free energy is calculated at the equilibrium volume for that pressure, and the Gibbs free energy follows by adding $pv$.

Starting from the absolute free energy at $T_0$, the free energy of the same phase over a continuous temperature interval is obtained using reversible scaling \cite{deKoning1999}.
The potential energy is scaled according to $U_{\lambda}=\lambda U$ with $\lambda=T_0/T$,
while the thermostat remains at $T_0$.
At zero pressure, scaling the potential in this way maps the system onto the unscaled potential at temperature $T=T_0/\lambda$.
The Gibbs free energy per atom along the temperature path is

\begin{equation}
\frac{g(T)}{T}
=
\frac{g(T_0)}{T_0}
+
\frac{3}{2}
k_{\mathrm{B}}
\ln
\left(
\frac{T_0}{T}
\right)
+
\frac{1}{2T_0}
\left[
w_{\mathrm{f}}(\lambda)
-
w_{\mathrm{b}}(\lambda)
\right],
\label{eq:rs}
\end{equation}

where $w_{\mathrm{f}}(\lambda)$ and $w_{\mathrm{b}}(\lambda)$ are the accumulated work per atom along the forward and backward scaling paths.
Their difference gives the reversible contribution in Eq.~\eqref{eq:rs}, while their sum provides the corresponding dissipation.

For each transition, the free energy branches of both phases are calculated over the same temperature interval.
The interval contains both the current estimate of the transformation temperature and the temperature targeted by the next correction, and is extended by 75~K on both sides.
The low temperature phase is initialized at the lower end of this interval and scaled to higher temperature.
The high temperature phase is initialized at the upper end and scaled to lower temperature.
The absolute free energy calculation is therefore performed on the stable side of each phase, while reversible scaling extends both branches across the transition region.
The transformation temperature is obtained from the crossing of the measured free energy branches, and the transition entropy is obtained from a linear fit of their difference within 60~K of the crossing.
For Ti and Zr melting, the reference temperature of the absolute free energy calculation of bcc is limited to $0.90\,T_A$.
Closer to the melting temperature, atoms in bcc Ti and Zr leave their lattice sites, which the Einstein crystal reference does not describe.

Finite pressure transformation temperatures are calculated with the same procedure in the isothermal isobaric ensemble at the target pressure $P$.
The absolute free energy at the target pressure is calculated at the equilibrium volume for that pressure, as described above.
For reversible scaling, the potential energy and the pressure are scaled together, $U_{\lambda}=\lambda U$ and $P_{\lambda}=\lambda P$.
The reversible scaling work is therefore accumulated from the generalized potential $U+PV$, $\mathrm{d}w=(U+PV)\,\mathrm{d}\lambda$.
This maps the calculation at temperature $T_0$ and pressure $P_{\lambda}$ to the unscaled system at temperature $T=T_0/\lambda$ and pressure $P$.
The phase crossing and transition entropy are then determined from the corresponding finite pressure Gibbs free energy branches.
At finite pressure, the branches cover fixed temperature windows around the reference transformation temperature, or, for GRACE, around its own transformation temperature, with 50~ps per scaling direction.

All free energy calculations use systems containing approximately $10^3$ atoms.
Each calculation begins with 25~ps of equilibration.
The absolute free energy calculation uses 50~ps for each switching direction.
The reversible scaling calculation uses 50~ps per 150~K of temperature interval for each scaling direction, with an additional 5~ps equilibration at the end of the forward sweep before the reverse sweep is started.


\subsection{Sampling of the fitting configurations}
\label{sec:sampling}

The configurations used to fit the correction layer are generated with isothermal isobaric molecular dynamics simulations.
At correction round $k$, configurations are sampled with the current potential $U_k$ defined in Sec.~\ref{sec:fitting}.
For $k=0$, the sampled potential is the unmodified base potential.

For each transition, both phases are sampled at the current estimate of the transformation temperature and at temperatures 75~K above and below it.
Two additional low temperature state points are included for the stable low temperature solid.
These are placed at 300~K and midway between 300~K and the lowest transition region temperature.
The additional state points extend the constraint on the correction away from the transition region and limit changes to the temperature dependence of the solid free energy.

Each state point is sampled independently using a system containing approximately 1000 atoms.
The simulations are performed with a time step of 1~fs for 25~ps.
Frames are written every 0.5~ps, and 30 configurations from each state point are used in the fit.
Liquid configurations are prepared by first melting the corresponding crystal and subsequently equilibrating at the target temperature.
We verify that no phase transitions occur in these trajectories using the Steinhardt bond orientational order parameter \cite{Steinhardt1983,Menon2019}.

All sampling calculations are performed at zero pressure, except for the pressure informed Ti correction layer.
For this layer, additional fitting configurations are generated at 1, 3, and 5~GPa using the same sampling protocol, with 14 state points at each pressure.
For hcp phases, an anisotropic barostat is used so that the cell shape and $c/a$ ratio can relax.
For cubic solids and liquids, an isotropic barostat is used.


\subsection{Correction fitting and validation}
\label{sec:fitmethods}

For most elements, the correction uses a linear ACE basis with 62 functions.
The basis contains 12 two body radial functions, 24 three body functions, and 26 four body functions.
The ACE basis is generated with \texttt{pacemaker} from the \texttt{python-ace} package \cite{Bochkarev2022}.
We use simplified spherical Bessel radial functions \cite{Bochkarev2022} and a cutoff of 6.5~\AA{}.
The correction layers are evaluated in LAMMPS using the ML-PACE implementation \cite{Lysogorskiy2021}.
The atomic energy is represented as a linear combination of the ACE basis functions, so that the correction is linear in its coefficients.
For Ag and Au, which require the largest thermodynamic corrections in the present set, larger ACE bases, 78 and 136 functions, respectively, are used to provide the additional flexibility required by the fit.

At correction step $k$, the accumulated coefficient vector is $\mathbf{c}_k$ and a new increment $\delta\mathbf{c}_k$ is fitted.
The thermodynamic targets apply to this increment, because they are constructed from the free energies of the potential currently deployed in the simulation.
The regularization and stability constraints act on the accumulated layer after the update.
To assign targets to individual configurations, one phase $p_0$ is chosen as the gauge phase,

\begin{equation}
r_{k,p_0}
=
0,
\qquad
r_{k,\alpha}
-
r_{k,\beta}
=
r_{\alpha\beta,k}.
\label{eq:phaseresidual}
\end{equation}

For systems containing a single transition, the solid is chosen as the gauge phase, so that the correction retains the crystalline energetics of the base model.
For systems containing both a solid to solid transition and melting, the phase shared by both transitions is chosen as the gauge phase.
This is the bcc phase for Ca, Ti, and Zr.
The ACE basis contains no constant term, so it cannot shift all phases uniformly, and the choice of gauge phase changes the fitted layer.
The correction energy of the gauge phase stays close to zero, and the free energy shift is applied to the other phases.
Averaged over the fitting configurations, the correction energy of the gauge phase is below 0.5~meV/atom for all elements.
The fit constrains energies and forces but not the stress, so the correction can change the volume of the gauge phase.

Each stored configuration $\mathbf{R}_i$ of phase $p$ from the current correction round contributes one row to the fit.
The row is the per atom basis vector $A_{i\nu}=a_\nu(\mathbf{R}_i)$ defined in Sec.~\ref{sec:fitting}, so that $\mathbf{A}_i\cdot\delta\mathbf{c}_k$ is the correction energy per atom predicted for configuration $i$.
To first order, the phase average of the correction increment must equal the phase target $r_{k,p}$.
Each configuration of phase $p$ is therefore assigned the fitting condition

\begin{equation}
\mathbf{A}_i
\cdot
\delta\mathbf{c}_k
=
r_{k,p}.
\label{eq:frametarget}
\end{equation}

The same first order residual $r_{k,p}$ is assigned to all configurations of phase $p$ sampled in the current round.
This includes configurations sampled at the measured crossing, the flanking temperatures, and the additional low temperature solid state points described in Sec.~\ref{sec:sampling}.
The use of a common target for each phase favors a rigid shift of the phase free energy over the sampled temperature range.

We additionally include one weighted row for each transition that directly constrains the difference between the two phase averages at the measured crossing of the current potential.
For a transition between phases $\alpha$ and $\beta$, we define

\begin{equation}
\Delta\mathbf{A}_{\alpha\beta,k}
=
\left\langle
\mathbf{A}
\right\rangle_{k,\alpha,T_k}
-
\left\langle
\mathbf{A}
\right\rangle_{k,\beta,T_k},
\label{eq:crossdesign}
\end{equation}

where $T_k$ is the measured transformation temperature of the current potential for this transition.
The transition row imposes

\begin{equation}
\Delta\mathbf{A}_{\alpha\beta,k}
\cdot
\delta\mathbf{c}_k
=
\eta_{\alpha\beta,k}
\left[
r_{k,\alpha}
-
r_{k,\beta}
\right].
\label{eq:crossrow}
\end{equation}

The averages in Eq.~\eqref{eq:crossdesign} are taken at the crossing measured for the potential deployed in the current round.
The transition row contains the same thermodynamic information as the individual configuration targets but places additional weight on the relative free energy difference that determines the transformation temperature.
It enters the objective with weight $10^2$, implemented by scaling the row and target by 10.
For Ag and Au the weight is $10^4$.
The factor $\eta_{\alpha\beta,k}$ damps the requested free energy shift and acts as a step length for the Newton update of the transformation temperature.
Solid to solid transitions have small transition entropies, so the first order estimate moves the crossing strongly for a small energy change and a full step tends to overshoot.
We use $\eta_{\alpha\beta,k}=1$ for melting transitions and $\eta_{\alpha\beta,k}=1/2$ for solid to solid transitions.
For melting transitions, $\eta_{\alpha\beta,k}$ is also set to $1/2$ when the deviations of the last two corrected potentials from $T_A$ have opposite signs and the last one lies outside the convergence tolerance.
The damping factor is applied to the transition row only, while the configuration rows in Eq.~\eqref{eq:frametarget} keep the full phase targets $r_{k,p}$.
The least squares solution is therefore a weighted compromise between the two, and the realized step is not fixed by $\eta_{\alpha\beta,k}$.
The requested and realized shifts of all rounds are listed in Table~S1.

The thermodynamic targets constrain the phase averaged correction energies but do not uniquely determine the ACE coefficients.
We therefore regularize both the magnitude of the correction and the forces introduced by it.
For a configuration $\mathbf{R}$, we define the force derivative matrix

\begin{equation}
D_{(j\mu),\nu}(\mathbf{R})
=
\frac{\partial}{\partial r_{j\mu}}
\sum_q
B_\nu(\sigma_q),
\label{eq:forcematrix}
\end{equation}

where $j$ labels the atom and $\mu$ the Cartesian component.
The force generated by the correction is $-\mathbf{D}\mathbf{c}$.
The corresponding force penalty matrix is

\begin{equation}
\mathbf{G}_F
=
\frac{1}{N_{\mathrm{fr}}}
\sum_i
\frac{
\mathbf{D}_i^{\mathsf T}
\mathbf{D}_i
}{
3N_i
},
\label{eq:forcepenalty}
\end{equation}

where $N_{\mathrm{fr}}$ is the number of fitting configurations.
The force penalty matrix is subsequently normalized to unit mean diagonal, so the prefactors in Eq.~\eqref{eq:forcepenalty} only define a convenient average before this normalization.

The thermodynamic configurations cover only a small part of configuration space.
We therefore impose additional stability constraints to prevent the correction from producing large energy changes in configurations that are poorly represented by the thermodynamic sampling data.
For this purpose, a random structure pool is generated using the ASSYST workflow by Poul et al. \cite{Poul2025}.

Per atom environments from the random structure pool are screened against the thermodynamic sampling data using an extrapolation grade $\gamma$ \cite{Podryabinkin2017,Lysogorskiy2023}.
Environments with $\gamma>2$ are insufficiently represented by the thermodynamic data and are included in the stability constraint pool.
The stability pool additionally contains dimer configurations spanning separations from 1.4 to 6.0~\AA{}.
The stability configurations are not assigned target energies.
Instead, the accumulated correction energy is constrained to remain within prescribed bounds, as defined in Eq.~\eqref{eq:fitobjective}.
Because the larger bases increase the flexibility of the correction, the Ag and Au fits include additional static configurations: ideal crystals along energy volume paths, the Bain path, and homogeneous shears of fcc.
They provide no thermodynamic targets and enter the stability constraints of Eq.~\eqref{eq:fitobjective} with bounds of $\pm2$~meV/atom on the accumulated correction, so that the 0~K energetics of the base model is kept for these structures.

The stability matrix $\mathbf{C}_f$ contains the corresponding ACE basis rows.
For random and harvested environments, these rows are single environment per atom basis vectors, so the bounds apply to the correction energy of individual atomic environments.
For dimer configurations, the rows are constructed for the total dimer correction energy.

After each update, the accumulated coefficient vector is $\mathbf{c}_{k+1}=\mathbf{c}_k+\delta\mathbf{c}_k$.
The coefficient increment is obtained by minimizing

\begin{equation}
\begin{aligned}
\min_{\delta\mathbf{c}_k}\quad
&
\left\|
\mathbf{A}\delta\mathbf{c}_k
-
\mathbf{y}_k
\right\|^2
+
\rho
\left\|
\delta\mathbf{c}_k
\right\|^2
+
\mu
\left\|
\mathbf{c}_{k+1}
\right\|^2
\\
&+
w_F
\mathbf{c}_{k+1}^{\mathsf T}
\mathbf{G}_F
\mathbf{c}_{k+1}
\\
\mathrm{s.t.}\quad
&
\mathbf{l}_f
\le
\mathbf{C}_f
\mathbf{c}_{k+1}
\le
\mathbf{u}_f.
\end{aligned}
\label{eq:fitobjective}
\end{equation}

Here, $\mathbf{A}$ and $\mathbf{y}_k$ contain the individual configuration rows together with the weighted transition rows, with energies in eV per atom.
The vectors $\mathbf{l}_f$ and $\mathbf{u}_f$ contain the lower and upper bounds for the stability rows.
For random and harvested environments, the bounds are $-0.05$ to $0.30$~eV per atom.
For dimer configurations, the bounds are $-0.15$ to $0.50$~eV per dimer.
The first term fits the thermodynamic targets.
The second term provides ridge regularization of the coefficient increment.
The third term limits the magnitude of the accumulated correction, while the fourth penalizes forces introduced by it.
Before solving Eq.~\eqref{eq:fitobjective}, the ACE basis columns are normalized by their root mean square magnitude over the fitting configurations.
The coefficient regularization and force penalty are evaluated in the same normalized representation, and the force penalty matrix is scaled to unit mean diagonal.
We use $\rho=10^{-6}$, $\mu=10^{-2}$, and $w_F=1$.
Equation~\eqref{eq:fitobjective} is solved using OSQP \cite{Stellato2020}.

The corrected potential is subsequently tested in short molecular dynamics probe simulations.
These include the liquid at 100~K above the highest sampled liquid temperature, each solid at 100~K above its highest valid sampled temperature, and each solid at 94\% of its equilibrium volume at an intermediate sampled temperature.
Each probe simulation is 5~ps long.
Atomic environments are harvested from these simulations if their extrapolation grade satisfies $\gamma>2$ relative to the stability envelope or if their per atom correction energy is below $-0.05$~eV.
Harvested environments are added to the stability pool with the same bounds as the random structure environments, and the correction is refitted.

The phase free energies of the resulting potential are then measured again as described in Sec.~\ref{sec:rs}.
Each round fits all transitions of an element together, with every transition aimed at its reference temperature, whether or not it is already within the tolerance.
The rounds stop when the measured transformation temperatures of all transitions of the element are within 15~K of the reference for melting and 50~K for solid to solid transitions.
A transition that is already within the tolerance therefore receives a small target, and its temperature can still change when the other transition of the element is corrected.
The final layers are measured with three independent reversible scaling calculations, and the reported transformation temperatures are given as mean and standard deviation.


\subsection{Validation of the corrected potentials}
\label{sec:validation}

We compare structural and thermodynamic properties of the base and the corrected potentials to test whether the correction degrades the description of the base model.
For Ag and Au, the static crystal configurations of Sec.~\ref{sec:fitmethods} enter the fit as constraints, so that their bulk moduli and energy volume curves stay close to those of the base model.
For each element, calculations are performed with the unmodified \texttt{GRACE-2L-SMAX-OMAT-large} potential and with the converged correction layer.

We first evaluate static properties of the solid phase that is stable at 0~K.
The structures are relaxed to zero pressure using anisotropic cell relaxation followed by conjugate gradient minimization.
Energy volume curves are calculated at 13 isotropic volume scales between 0.94 and 1.06 of the relaxed volume.
The resulting energies are fitted to a third order Birch--Murnaghan equation of state to obtain the bulk modulus.

Energy volume curves are also calculated for all polymorphs of each element listed in the Materials Project~\cite{Jain2013}.
The fcc, bcc, hcp, dhcp, $\omega$, simple cubic, diamond, and A15 structures are added when they are not present in the database.
Each structure is relaxed with its cell shape fixed, and its energy is then calculated at 13 volumes between 0.94 and 1.06 of the relaxed volume.
Structures that change symmetry during unconstrained relaxation are excluded from the comparison.

Vacancy formation energies are calculated in supercells of the ground state structure containing 500 atoms for fcc, 686 atoms for bcc, and 384 atoms for hcp.
The supercell containing the vacancy is first relaxed at the volume of the perfect crystal and then relaxed again at zero pressure.
The vacancy formation energy is $E_{f}=E_{N-1}-(N-1)E_{N}/N$,
where $E_{N}$ and $E_{N-1}$ are the energies of the relaxed perfect and defective supercells at zero pressure.

The temperature dependence of the solid volume is calculated in the isothermal isobaric ensemble at zero pressure using supercells containing about 1000 atoms.
Nine temperatures are sampled between 300~K and $0.9\,T_{\mathrm{m}}$.
For the corrected potential, $T_{\mathrm{m}}$ is the reference melting temperature.
For the base model, $T_{\mathrm{m}}$ is the melting temperature calculated for \texttt{GRACE-2L-SMAX-OMAT-large}.
Each temperature is equilibrated for 20~ps followed by 40~ps of sampling, from which the average volume per atom is obtained.
For hcp phases, the cell axes are allowed to relax independently.
Temperatures at which the crystal melts or transforms to another structure are excluded.
These cases are identified from the solid fraction and common neighbor analysis of the final configuration.
The mean linear thermal expansion coefficient, $\bar{\alpha}_L=[L(T_{\mathrm{hi}})-L(293~\mathrm{K})]/[L(293~\mathrm{K})(T_{\mathrm{hi}}-293~\mathrm{K})]$ with $L=V^{1/3}$, is evaluated from 293~K to an upper temperature $T_{\mathrm{hi}}$ between 748 and 1100~K that lies within the experimental data and up to which the solid keeps its crystal structure in the simulations of both potentials.
Since the simulations are classical, the calculated $\alpha_L$ does not include the reduction of thermal expansion by quantum effects below the Debye temperature.

The liquid density is calculated at the melting point of each potential in the isothermal isobaric ensemble at zero pressure using supercells containing about 1000 atoms.
The liquid is generated by melting the crystal at $1.5\,T_{\mathrm{m}}$ for 10~ps.
This is followed by 20~ps of equilibration and 40~ps of sampling at $T_{\mathrm{m}}$.
The absence of crystalline order is confirmed from the solid fraction of the final configuration.

The liquid structure is evaluated from radial distribution functions at $1.03\,T_A$.
The simulations are performed in the isothermal isobaric ensemble at zero pressure using 10~ps of equilibration followed by 30~ps of sampling.
Configurations are stored every 0.5~ps, and the radial distribution function is averaged over the second half of the trajectory using a bin width of 0.05~\AA{}.
The liquid densities at the common temperature $1.03\,T_A$ are taken from the same simulations.

The experimental bulk moduli are obtained from single crystal elastic constants~\cite{Neighbours1958,Kamm1964,Overton1955,Slutsky1957,Featherston1963,Alers1960Ni,Fisher1964,Alers1960V} and, for Ca, from the equation of state~\cite{Anzellini2018}.
The experimental thermal expansion coefficients are taken from Ref.~\cite{Touloukian1975}, the vacancy formation energies from positron annihilation measurements of the mean lifetime, Doppler broadening, or angular correlation~\cite{Schaefer1987}, and the liquid densities from Refs.~\cite{Abdullaev2020,Paradis2008,Assael2006,Assael2010,Abdullaev2015,Abdullaev2019,Paradis2002,Paradis2000,Ishikawa2005}.

\section*{Data availability}

Fitted correction layers are available in \cite{GraceCorr2026}.

\section*{Code availability}

All simulations were performed with publicly available software.
The code for the constrained fit and the iteration procedure will be released after publication.
The \texttt{GRACE-2L-SMAX-OMAT-large} potential was evaluated using \texttt{grace-tensorpotential} (\url{https://github.com/ICAMS/grace-tensorpotential}) together with LAMMPS \cite{Thompson2022}.
Free energy calculations and reversible scaling were performed using \texttt{calphy} (\url{https://github.com/ICAMS/calphy}).
The ACE correction layers were constructed using \texttt{python-ace} (\url{https://github.com/ICAMS/python-ace}) and evaluated in LAMMPS using the ML-PACE implementation.
Random structures used for the stability constraints were generated with ASSYST (\url{https://github.com/eisenforschung/assyst}) and PyXtal \cite{Fredericks2021}.
Reference thermodynamic quantities from SGTE were evaluated using pycalphad \cite{Otis2017}.

\section*{Author contributions}

S.M. and R.D. contributed to conceptualization and methodology.
S.M. developed the software.
S.M. and B.S.O. performed the investigation.
S.M. carried out the formal analysis, validation, and visualization.
S.M., B.S.O., and R.D. contributed to writing the original draft and to review and editing of the manuscript.

\section*{Competing interests}

The authors declare no competing interests.

\section*{Acknowledgments}

Calculations for this publication were performed on the HPC cluster Elysium of the Ruhr University Bochum, subsidized by the DFG (INST 213/1055-1).

\bibliographystyle{simplejournal}
\bibliography{references}

\end{document}


\maketitle

\section{Additional validation of the corrected potentials}

The main text compares selected structural and thermodynamic properties of the base GRACE model and the corrected model.
Here we provide the full validation data used to assess changes in static phase stability, finite temperature solid volumes, and liquid structure.
All calculations are performed using the protocols described in the Methods section of the main text, under Validation of the corrected potentials.
The dotted curves correspond to the unmodified \texttt{GRACE-2L-SMAX-OMAT-large} potential.
The solid curves correspond to the same base model with the converged correction layer.

\begin{figure*}[htp!]
    \centering
    \includegraphics[width=1.0\linewidth]{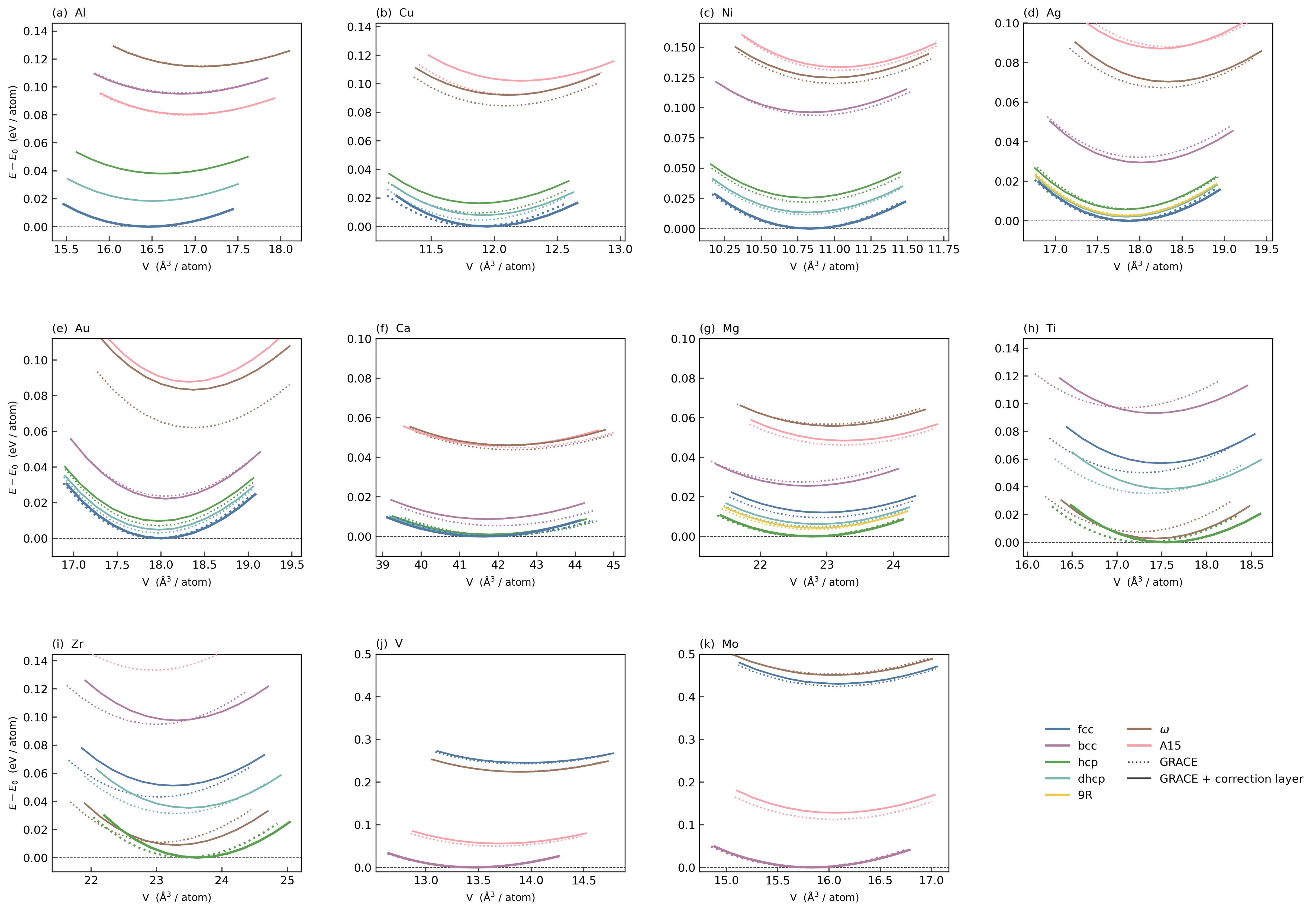}
    \caption{
    Static energy volume curves for polymorphs of the elements considered in this work.
    Energies are plotted relative to the minimum energy of the experimental ground state structure calculated with the same potential.
    Dotted curves show the unmodified GRACE model, and solid curves show GRACE with the correction layer.
    The correction preserves smooth energy volume curves and does not change the zero temperature phase ordering for the relevant structures.
    }
    \label{fig:si_polymorph_ev}
\end{figure*}

Figure~\ref{fig:si_polymorph_ev} compares static energy volume curves for the polymorphs included in the validation set.
The correction changes the relative energies only weakly on the scale shown for most structures.
The minima remain smooth after correction, and the low energy phase ordering is retained.
This supports the interpretation that the correction acts mainly as a thermodynamic adjustment of relative phase free energies rather than as a replacement of the static energy landscape of the base model.

\begin{figure*}[htp!]
    \centering
    \includegraphics[width=1.0\linewidth]{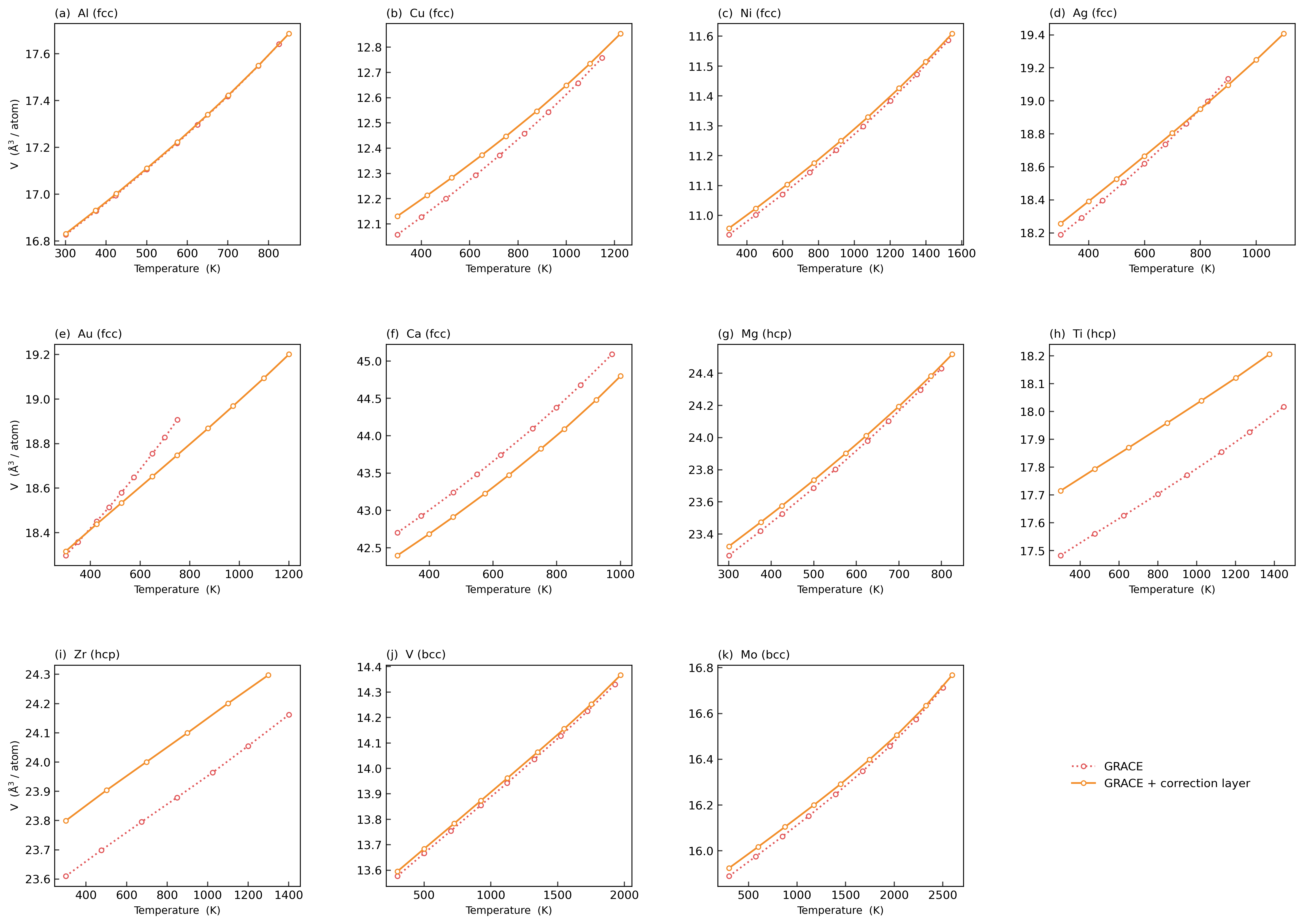}
    \caption{
    Solid volume as a function of temperature for the structures used in the thermal expansion analysis, as indicated in each panel.
    Dotted curves show the unmodified GRACE model, and solid curves show GRACE with the correction layer.
    Temperatures at which the solid melted or transformed to another structure are excluded.
    The corrected curves remain close to the base model for most systems and remain smooth over the sampled temperature range.
    }
    \label{fig:si_volume_temperature}
\end{figure*}

Figure~\ref{fig:si_volume_temperature} shows the finite temperature volume curves used to obtain the linear thermal expansion coefficients reported in the main text.
The correction preserves the temperature dependence of the solid volume for most elements.
Larger shifts are visible for Ti, Zr, and Ca, where the correction changes the volume at 300~K by $+1.3$\%, $+0.8$\%, and $-0.7$\%, respectively.
In these elements the phase at 300~K is not the gauge phase and carries the correction of the solid to solid transition.
For Cu, whose fcc phase is the gauge phase, the volume at 300~K changes by $+0.6$\%.
The slopes of the volume curves remain close to those of the base model, except for Au, Ag, and Cu, where the correction reduces the thermal expansion.

\begin{figure*}[htp!]
    \centering
    \includegraphics[width=1.0\linewidth]{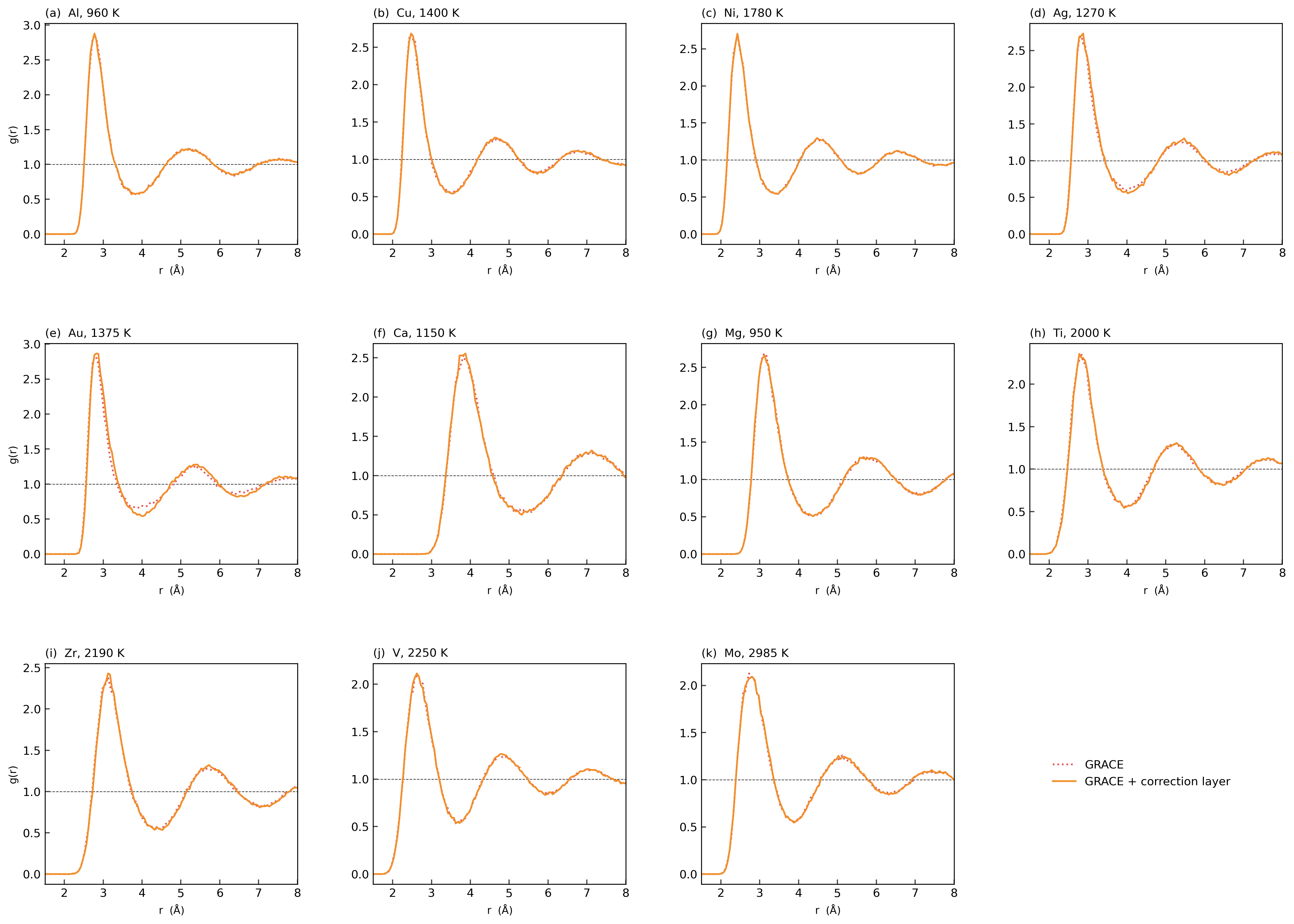}
    \caption{
    Liquid radial distribution functions at $1.03\,T_A$, where $T_A$ is the SGTE melting temperature.
    Dotted curves show the unmodified GRACE model, and solid curves show GRACE with the correction layer.
    The local liquid structure is largely preserved after correction for all elements considered here.
    }
    \label{fig:si_liquid_gr}
\end{figure*}

Figure~\ref{fig:si_liquid_gr} compares liquid radial distribution functions before and after applying the correction layer.
The positions and heights of the main coordination peaks remain close to those of the base model.
The correction therefore modifies the relative solid and liquid free energies without introducing large changes in the local liquid structure.

\section{Correction rounds}

\renewcommand{\thetable}{S\arabic{table}}
\begin{table*}[htp!]
    \caption{
    Record of every correction round.
    Round $k$ fits the increment from $U_k$ to $U_{k+1}$, where $U_0$ is the base model.
    $T_k$ is the transformation temperature of $U_k$ at which the round is sampled, and $\Delta s_k=s_\alpha-s_\beta$ is the transition entropy used to compute the free energy shift $r_k=\Delta s_k(T_A-T_k)$ of Eq.~2.
    $\eta_k$ is the damping of the transition row.
    The realized shift is the change of $\langle \Delta U\rangle_\alpha-\langle \Delta U\rangle_\beta$ produced by the fitted increment at $T_k$.
    Shifts are given in meV/atom.
    $T_{k+1}-T_A$ is obtained from a single reversible scaling sweep.
    Dashes mark crossings that were not measured, so round 0 of Ca fcc$\to$bcc requested no shift.
    }
    \label{tab:si_rounds}
    \centering
\vspace{6pt}
\footnotesize
\setlength{\tabcolsep}{5pt}
\begin{tabular}{llrrrrrrrrr}
\toprule
Element & Transition & $k$ & $T_k$ & $T_k-T_A$ & $\Delta s_k$ & $r_k$ & $\eta_k$ & $\eta_k r_k$ & realized & $T_{k+1}-T_A$ \\
 & & & (K) & (K) & (J\,mol$^{-1}$\,K$^{-1}$) & (meV) & & (meV) & (meV) & (K) \\
\midrule
Al & melting & 0 & 926 & $-7$ & $-11.28$ & $-0.84$ & 1.0 & $-0.84$ & $-0.83$ & $-2$ \\
\addlinespace
Cu & melting & 0 & 1270 & $-88$ & $-9.49$ & $-8.61$ & 1.0 & $-8.61$ & $-8.43$ & $+3$ \\
\addlinespace
Ni & melting & 0 & 1681 & $-47$ & $-10.48$ & $-5.11$ & 1.0 & $-5.11$ & $-5.04$ & $+4$ \\
\addlinespace
Ag & melting & 0 & 991 & $-244$ & $-9.59$ & $-24.29$ & 1.0 & $-24.29$ & $-24.29$ & $-8$ \\
\addlinespace
Au & melting & 0 & 842 & $-496$ & $-9.32$ & $-47.86$ & 1.0 & $-47.86$ & $-47.83$ & $-42$ \\
 &  & 1 & 1295 & $-42$ & $-8.35$ & $-3.67$ & 1.0 & $-3.67$ & $-3.63$ & $+15$ \\
\addlinespace
Mg & melting & 0 & 899 & $-24$ & $-9.29$ & $-2.30$ & 1.0 & $-2.30$ & $-2.24$ & $+4$ \\
\addlinespace
Ca & fcc$\to$bcc & 0 & -- & -- & -- & $0.00$ & 0.5 & $0.00$ & $+0.08$ & -- \\
 &  & 1 & 537 & $-179$ & $-1.30$ & $-2.41$ & 0.5 & $-1.20$ & $-1.56$ & $-118$ \\
 &  & 2 & 598 & $-118$ & $-1.16$ & $-1.41$ & 0.5 & $-0.71$ & $-0.89$ & $-12$ \\
 & melting & 0 & 1078 & $-37$ & $-7.49$ & $-2.88$ & 1.0 & $-2.88$ & $-2.70$ & $+24$ \\
 &  & 1 & 1139 & $+24$ & $-7.31$ & $+1.82$ & 1.0 & $+1.82$ & $+1.83$ & $+10$ \\
 &  & 2 & 1125 & $+10$ & $-7.12$ & $+0.77$ & 1.0 & $+0.77$ & $+0.75$ & $-15$ \\
\addlinespace
Ti & hcp$\to$bcc & 0 & 1283 & $+128$ & $-2.83$ & $+3.76$ & 0.5 & $+1.88$ & $+2.89$ & $+35$ \\
 &  & 1 & 1190 & $+35$ & $-3.12$ & $+1.13$ & 0.5 & $+0.57$ & $+0.91$ & $+1$ \\
 & melting & 0 & 1788 & $-153$ & $-6.37$ & $-10.09$ & 1.0 & $-10.09$ & $-9.37$ & $-18$ \\
 &  & 1 & 1923 & $-18$ & $-6.74$ & $-1.28$ & 1.0 & $-1.28$ & $-0.95$ & $-4$ \\
\addlinespace
Zr & hcp$\to$bcc & 0 & 1130 & $-9$ & $-1.92$ & $-0.19$ & 0.5 & $-0.09$ & $+0.21$ & $-36$ \\
 & melting & 0 & 1947 & $-181$ & $-6.51$ & $-12.18$ & 1.0 & $-12.18$ & $-11.61$ & $+5$ \\
\addlinespace
V & melting & 0 & 2149 & $-34$ & $-10.42$ & $-3.64$ & 1.0 & $-3.64$ & $-3.60$ & $+8$ \\
\addlinespace
Mo & melting & 0 & 2791 & $-105$ & $-13.22$ & $-14.41$ & 1.0 & $-14.41$ & $-14.31$ & $-8$ \\
\bottomrule
\end{tabular}
\end{table*}

\section{Liquid density at a common temperature}
\begin{table}[htp!]
    \caption{
    Liquid density of GRACE and of GRACE with the correction layer at $1.03\,T_A$, from the simulations of Fig.~\ref{fig:si_liquid_gr}.
    }
    \label{tab:si_density}
    \centering
\vspace{6pt}
\begin{tabular}{lrrrr}
\toprule
Element & $T$ & \multicolumn{2}{c}{$\rho$ (g\,cm$^{-3}$)} & change \\
\cmidrule(lr){3-4}
 & (K) & GRACE & corrected & (\%) \\
\midrule
Al & 960 & 2.328 & 2.329 & $+0.1$ \\
Cu & 1400 & 7.644 & 7.675 & $+0.4$ \\
Ni & 1780 & 7.819 & 7.829 & $+0.1$ \\
Ag & 1270 & 8.416 & 8.665 & $+3.0$ \\
Au & 1375 & 15.243 & 16.121 & $+5.8$ \\
Mg & 950 & 1.532 & 1.537 & $+0.3$ \\
Ca & 1150 & 1.392 & 1.405 & $+1.0$ \\
Ti & 2000 & 4.262 & 4.225 & $-0.9$ \\
Zr & 2190 & 6.030 & 6.005 & $-0.4$ \\
V & 2250 & 5.616 & 5.614 & $0.0$ \\
Mo & 2985 & 8.844 & 8.865 & $+0.2$ \\
\bottomrule
\end{tabular}
\end{table}